\documentclass[reqno,12pt,a4paper]{amsart}

\usepackage{mathrsfs}
\usepackage{amssymb}
\usepackage{amsfonts}
\usepackage{latexsym}
\usepackage{amsthm}

\usepackage{graphicx}
\def\lb{\label}

\newcommand{\er}[1]{\textrm{(\ref{#1})}}

\begin{document}

%%%%%%%%%% Some definitions %%%%%%%%%%

%%%%%%%% Equations, theorems %%%%%%%%%
\renewcommand{\theequation}{\arabic{section}.\arabic{equation}}
\theoremstyle{plain}
\newtheorem{theorem}{\bf Theorem}[section]
\newtheorem{lemma}[theorem]{\bf Lemma}
\newtheorem{corollary}[theorem]{\bf Corollary}
\newtheorem{proposition}[theorem]{\bf Proposition}
\newtheorem{definition}[theorem]{\bf Definition}
\newtheorem{remark}[theorem]{\it Remark}
%\theoremstyle{remark}
%\newtheorem{remark}[theorem]{\bf Remark}

%%%%% Alphabet %%%%%
\def\a{\alpha}  \def\cA{{\mathcal A}}     \def\bA{{\bf A}}  \def\mA{{\mathscr A}}
\def\b{\beta}   \def\cB{{\mathcal B}}     \def\bB{{\bf B}}  \def\mB{{\mathscr B}}
\def\g{\gamma}  \def\cC{{\mathcal C}}     \def\bC{{\bf C}}  \def\mC{{\mathscr C}}
\def\G{\Gamma}  \def\cD{{\mathcal D}}     \def\bD{{\bf D}}  \def\mD{{\mathscr D}}
\def\d{\delta}  \def\cE{{\mathcal E}}     \def\bE{{\bf E}}  \def\mE{{\mathscr E}}
\def\D{\Delta}  \def\cF{{\mathcal F}}     \def\bF{{\bf F}}  \def\mF{{\mathscr F}}
\def\c{\chi}    \def\cG{{\mathcal G}}     \def\bG{{\bf G}}  \def\mG{{\mathscr G}}
\def\z{\zeta}   \def\cH{{\mathcal H}}     \def\bH{{\bf H}}  \def\mH{{\mathscr H}}
\def\e{\eta}    \def\cI{{\mathcal I}}     \def\bI{{\bf I}}  \def\mI{{\mathscr I}}
\def\p{\psi}    \def\cJ{{\mathcal J}}     \def\bJ{{\bf J}}  \def\mJ{{\mathscr J}}
\def\vT{\Theta} \def\cK{{\mathcal K}}     \def\bK{{\bf K}}  \def\mK{{\mathscr K}}
\def\k{\kappa}  \def\cL{{\mathcal L}}     \def\bL{{\bf L}}  \def\mL{{\mathscr L}}
\def\l{\lambda} \def\cM{{\mathcal M}}     \def\bM{{\bf M}}  \def\mM{{\mathscr M}}
\def\L{\Lambda} \def\cN{{\mathcal N}}     \def\bN{{\bf N}}  \def\mN{{\mathscr N}}
\def\m{\mu}     \def\cO{{\mathcal O}}     \def\bO{{\bf O}}  \def\mO{{\mathscr O}}
\def\n{\nu}     \def\cP{{\mathcal P}}     \def\bP{{\bf P}}  \def\mP{{\mathscr P}}
\def\r{\rho}    \def\cQ{{\mathcal Q}}     \def\bQ{{\bf Q}}  \def\mQ{{\mathscr Q}}
\def\s{\sigma}  \def\cR{{\mathcal R}}     \def\bR{{\bf R}}  \def\mR{{\mathscr R}}
\def\S{\Sigma}  \def\cS{{\mathcal S}}     \def\bS{{\bf S}}  \def\mS{{\mathscr S}}
\def\t{\tau}    \def\cT{{\mathcal T}}     \def\bT{{\bf T}}  \def\mT{{\mathscr T}}
\def\f{\phi}    \def\cU{{\mathcal U}}     \def\bU{{\bf U}}  \def\mU{{\mathscr U}}
\def\F{\Phi}    \def\cV{{\mathcal V}}     \def\bV{{\bf V}}  \def\mV{{\mathscr V}}
\def\P{\Psi}    \def\cW{{\mathcal W}}     \def\bW{{\bf W}}  \def\mW{{\mathscr W}}
\def\o{\omega}  \def\cX{{\mathcal X}}     \def\bX{{\bf X}}  \def\mX{{\mathscr X}}
\def\x{\xi}     \def\cY{{\mathcal Y}}     \def\bY{{\bf Y}}  \def\mY{{\mathscr Y}}
\def\X{\Xi}     \def\cZ{{\mathcal Z}}     \def\bZ{{\bf Z}}  \def\mZ{{\mathscr Z}}
\def\O{\Omega}

%*********************
\def\be{{\bf e}}  \def\bp{{\bf p}} \def\bq{{\bf q}}  \def\br{{\bf r}}
\def\bv{{\bf v}} \def\bu{{\bf u}}
\def\Om{\Omega}
%************************
\def\bbD{\pmb \Delta}
\def\mm{\mathrm m}
\def\mn{\mathrm n}
%*************************

\newcommand{\mc}{\mathscr {c}}

\newcommand{\gA}{\mathfrak{A}}          \newcommand{\ga}{\mathfrak{a}}
\newcommand{\gB}{\mathfrak{B}}          \newcommand{\gb}{\mathfrak{b}}
\newcommand{\gC}{\mathfrak{C}}          \newcommand{\gc}{\mathfrak{c}}
\newcommand{\gD}{\mathfrak{D}}          \newcommand{\gd}{\mathfrak{d}}
\newcommand{\gE}{\mathfrak{E}}
\newcommand{\gF}{\mathfrak{F}}           \newcommand{\gf}{\mathfrak{f}}
\newcommand{\gG}{\mathfrak{G}}           %\newcommand{\gg}{\mathfrak{g}}
\newcommand{\gH}{\mathfrak{H}}           \newcommand{\gh}{\mathfrak{h}}
\newcommand{\gI}{\mathfrak{I}}           \newcommand{\gi}{\mathfrak{i}}
\newcommand{\gJ}{\mathfrak{J}}           \newcommand{\gj}{\mathfrak{j}}
\newcommand{\gK}{\mathfrak{K}}            \newcommand{\gk}{\mathfrak{k}}
\newcommand{\gL}{\mathfrak{L}}            \newcommand{\gl}{\mathfrak{l}}
\newcommand{\gM}{\mathfrak{M}}            \newcommand{\gm}{\mathfrak{m}}
\newcommand{\gN}{\mathfrak{N}}            \newcommand{\gn}{\mathfrak{n}}
\newcommand{\gO}{\mathfrak{O}}
\newcommand{\gP}{\mathfrak{P}}             \newcommand{\gp}{\mathfrak{p}}
\newcommand{\gQ}{\mathfrak{Q}}             \newcommand{\gq}{\mathfrak{q}}
\newcommand{\gR}{\mathfrak{R}}             \newcommand{\gr}{\mathfrak{r}}
\newcommand{\gS}{\mathfrak{S}}              \newcommand{\gs}{\mathfrak{s}}
\newcommand{\gT}{\mathfrak{T}}             \newcommand{\gt}{\mathfrak{t}}
\newcommand{\gU}{\mathfrak{U}}             \newcommand{\gu}{\mathfrak{u}}
\newcommand{\gV}{\mathfrak{V}}             \newcommand{\gv}{\mathfrak{v}}
\newcommand{\gW}{\mathfrak{W}}             \newcommand{\gw}{\mathfrak{w}}
\newcommand{\gX}{\mathfrak{X}}               \newcommand{\gx}{\mathfrak{x}}
\newcommand{\gY}{\mathfrak{Y}}              \newcommand{\gy}{\mathfrak{y}}
\newcommand{\gZ}{\mathfrak{Z}}             \newcommand{\gz}{\mathfrak{z}}

\def\ve{\varepsilon} \def\vt{\vartheta} \def\vp{\varphi}  \def\vk{\varkappa}
\def\vr{\varrho} \def\vs{\varsigma}

\def\A{{\mathbb A}} \def\B{{\mathbb B}} \def\C{{\mathbb C}}
\def\dD{{\mathbb D}} \def\E{{\mathbb E}} \def\dF{{\mathbb F}} \def\dG{{\mathbb G}}
\def\H{{\mathbb H}}\def\I{{\mathbb I}} \def\J{{\mathbb J}} \def\K{{\mathbb K}}
\def\dL{{\mathbb L}}\def\M{{\mathbb M}} \def\N{{\mathbb N}} \def\dO{{\mathbb O}}
\def\dP{{\mathbb P}} \def\dQ{{\mathbb Q}} \def\R{{\mathbb R}}\def\S{{\mathbb S}} \def\T{{\mathbb T}}
\def\U{{\mathbb U}} \def\V{{\mathbb V}}\def\W{{\mathbb W}} \def\X{{\mathbb X}}
\def\Y{{\mathbb Y}} \def\Z{{\mathbb Z}}

\def\dk{{\Bbbk}}

%%%%% Arrows %%%%%

\def\la{\leftarrow}              \def\ra{\rightarrow}            \def\Ra{\Rightarrow}
\def\ua{\uparrow}                \def\da{\downarrow}
\def\lra{\leftrightarrow}        \def\Lra{\Leftrightarrow}

%%%%% Typography %%%%%

\def\lt{\biggl}                  \def\rt{\biggr}
\def\ol{\overline}               \def\wt{\widetilde}
\def\no{\noindent}

%%%%% Math signs %%%%%

\let\ge\geqslant                 \let\le\leqslant
\def\lan{\langle}                \def\ran{\rangle}
\def\/{\over}                    \def\iy{\infty}
\def\sm{\setminus}               \def\es{\emptyset}
\def\ss{\subset}                 \def\ts{\times}
\def\pa{\partial}                \def\os{\oplus}
\def\om{\ominus}                 \def\ev{\equiv}
\def\iint{\int\!\!\!\int}        \def\iintt{\mathop{\int\!\!\int\!\!\dots\!\!\int}\limits}
\def\el2{\ell^{\,2}}             \def\1{1\!\!1}
\def\sh{\sharp}
\def\wh{\widehat}
\def\bs{\backslash}
\def\intl{\int\limits}
%%%%% Math operations %%%%%

\def\na{\mathop{\mathrm{\nabla}}\nolimits}
\def\sh{\mathop{\mathrm{sh}}\nolimits}
\def\ch{\mathop{\mathrm{ch}}\nolimits}
\def\where{\mathop{\mathrm{where}}\nolimits}
\def\all{\mathop{\mathrm{all}}\nolimits}
\def\as{\mathop{\mathrm{as}}\nolimits}
\def\Area{\mathop{\mathrm{Area}}\nolimits}
\def\arg{\mathop{\mathrm{arg}}\nolimits}
\def\const{\mathop{\mathrm{const}}\nolimits}
\def\det{\mathop{\mathrm{det}}\nolimits}
\def\diag{\mathop{\mathrm{diag}}\nolimits}
\def\diam{\mathop{\mathrm{diam}}\nolimits}
\def\dim{\mathop{\mathrm{dim}}\nolimits}
\def\dist{\mathop{\mathrm{dist}}\nolimits}
\def\Im{\mathop{\mathrm{Im}}\nolimits}
\def\Iso{\mathop{\mathrm{Iso}}\nolimits}
\def\Ker{\mathop{\mathrm{Ker}}\nolimits}
\def\Lip{\mathop{\mathrm{Lip}}\nolimits}
\def\rank{\mathop{\mathrm{rank}}\limits}
\def\Ran{\mathop{\mathrm{Ran}}\nolimits}
\def\Re{\mathop{\mathrm{Re}}\nolimits}
\def\Res{\mathop{\mathrm{Res}}\nolimits}
\def\res{\mathop{\mathrm{res}}\limits}
\def\sign{\mathop{\mathrm{sign}}\nolimits}
\def\span{\mathop{\mathrm{span}}\nolimits}
\def\supp{\mathop{\mathrm{supp}}\nolimits}
\def\Tr{\mathop{\mathrm{Tr}}\nolimits}
\def\BBox{\hspace{1mm}\vrule height6pt width5.5pt depth0pt \hspace{6pt}}
\def\where{\mathop{\mathrm{where}}\nolimits}
\def\as{\mathop{\mathrm{as}}\nolimits}

%%%%%%%%%%%%% specialities %%%%%%%%%%%%%%

\newcommand\nh[2]{\widehat{#1}\vphantom{#1}^{(#2)}}
%{{\mathop{#1}\limits^\wedge}\vphantom{#1}^{(#2)}}
\def\dia{\diamond}

\def\Oplus{\bigoplus\nolimits}

%%%%%%%%%%% End of definitions %%%%%%%%%%

%%%%% OLD OLD OLD

\def\qqq{\qquad}
\def\qq{\quad}
\let\ge\geqslant
\let\le\leqslant
\let\geq\geqslant
\let\leq\leqslant
\newcommand{\ca}{\begin{cases}}
\newcommand{\ac}{\end{cases}}
\newcommand{\ma}{\begin{pmatrix}}
\newcommand{\am}{\end{pmatrix}}
\renewcommand{\[}{\begin{equation}}
\renewcommand{\]}{\end{equation}}
\def\eq{\begin{equation}}
\def\qe{\end{equation}}
\def\[{\begin{equation}}
\def\bu{\bullet}
\def\ced{\centerdot}
\def\tes{\textstyle}

%\bigskip
%\medskip
%\smallskip

\title[{Spectral theory for periodic vector NLS equations
 }]
{Spectral  theory for periodic vector NLS equations}

\date{\today}

\author[Evgeny Korotyaev]{Evgeny Korotyaev}
\address{Department of Analysis,
Saint-Petersburg State University,   Universitetskaya nab. 7/9, St.
Petersburg, 199034, Russia, \ korotyaev@gmail.com, \
e.korotyaev@spbu.ru}


\begin{abstract}
\no We consider a first order operator with a periodic 3x3 matrix
potential on the real line. This operator appears in the problem of
the periodic vector NLS equation.  The spectrum of
the operator covers the real line, it is union of the spectral bands of
multiplicity 3, separated by spectral intervals of multiplicity 1.
The main results of this work are the following:
\\
$\cdot$   The Lyapunov function
on the corresponding 2 or 3-sheeted Riemann surface is described.
\\
$ \cdot$ Necessary and sufficient conditions are given when the Riemann surface is 2-sheeted.
\\
$\cdot$ The asymptotics of 2-periodic eigenvalues
are determined.
\\
$\cdot$ One constructs an entire function, which is positive on the
spectrum of multiplicity 3 and is negative on its gaps.
\\
$\cdot$  The Borg type results about inverse problems are solved.

\end{abstract}

\maketitle

\hfill 	{\it  To the memory of Pavel Kargaev
(1953-2024)}

%{\it \footnotesize Dedicated to the memory of Pavel Kargaev
%(1953-2024)}

\begin{quotation}
\begin{center}
{\bf Table of Contents}
\end{center}

\vskip 6pt

{\footnotesize

1. Introduction and main results \hfill \pageref{Sec1}\ \ \ \ \

2. Monodromy matrices and their properties  \hfill \pageref{Sec2}\ \ \ \ \

3. Traces of monodromy matrices and their properties  \hfill \pageref{Sec2x}\ \ \ \ \

4. Lyapunov functions \hfill \pageref{Sec3}\ \ \ \ \

5. Conformal mappings and trace formulas  \hfill \pageref{Sec4}\ \ \ \ \

6.  Appendix \hfill \pageref{Sec5}\ \ \ \ \

 }
\end{quotation}

%  \vskip 0.25cm

\section {Introduction and main results \lb{Sec1}}
\setcounter{equation}{0}

\subsection{Vector Nonlinear Schr\"odinger equations}
Consider the periodic 2 dimensional defocusing Vector Nonlinear Schr\"odinger equation (or the Manakov systems, or shortly the MS)
\[
\lb{VSE1}
\begin{aligned}
iv_t=-v_{xx}+ 2|v|^2v,\qqq v=(v_1,v_2)^\top \in\C^2,
\end{aligned}
\]
where $|v|^2=|v_1|^2+|v_2|^2$. Here $v_1, v_2$ represents as a sum of right- and left-hand polarized waves. The  NLS equation is one
of the most fundamental and the most universal nonlinear PDE.
Zakharov and Shabat proved that it   is integrable \cite{ZS72}.
The vector NLS equation appears in physics to study two mode optical fibers, photorefractive materials and so on. Manakov \cite{Ma74} obtained first basis results about the vector case and proved that it is completely integrable. His proof is based on the Lax representation.
There is a  number articles dedicated to periodic and quasi-periodic
solutions of  the MS, based on an algebrogeometric
technique, see e.g. Krichever \cite{Kr77}, Dubrovin \cite{D83}, Adams, Harnad, Hurtubise \cite{AHH90}, Elgin, Enolski, Its \cite{EEI07}, Warren,  Elgin
\cite{WE07}, where the inverse problem is discussed. The results
are based on the finite bands integration technique for the scalar case
\cite{DN74}, \cite{IM75},  \cite{N74}. They do not discuss the spectral properties
of $\cL$ and the existence of finite bands potentials is  still open problem.
The case of two bands is discussed by  Woodcock, Warren,  Elgin \cite{WE07}.
Results about the spectral theory for periodic vector NLS systems  are obtained in the  paper \cite{K10} devoted to the more general class of operators  and in particulr, $v$ is a matrix-valued function.

Scalar  NLS equations governs
the long-distance pulse propagation in optical fibers. Early
fundamental research on NLS equations appears in the papers of
Zakharov and Shabat \cite{ZS72}, who proved that the scalar NLS equation
is integrable.
For the case of the scalar NLS equation we consider the Zakharov-Shabat (shortly ZS) operators. Briefly describe results about the ZS-operator $\cL_{zs}$ on $L^2(\R,\C^2)$ given by
\[
\lb{ddzs} \tes
\cL_{zs}=iJ_{zs}{d\/ dx}+ V_{zs},\ \ \ \  J_{zs} =\ma 1&0\\ 0&-1\am, \
\ \ \ \ V_{zs}= \ma 0 & \ol u\\  u & 0\am,\qq u\in L^2(\T),
\]
see, e.g., \cite{LS91}.
The spectrum of $\cL_{zs}$ is purely absolutely
continuous and consists of spectral bands separated by the gaps:
\[
\lb{ddzs1}
\s(\cL_{zs})=\R\sm \cup \g_n, \qq \g_n=(\l_n^-,\l_n^+),\qq
\l_n^\pm=\pi n+o(1)\ \ \as \ n\to \pm \iy.
\]
Here $\l_n^\pm, n\in \Z$ are 2-periodic eigenvalues of the equation
$iJ_{zs}y'+ V_{zs}y=\l y$.
The corresponding Riemann surface is two-sheeted with the real branch points
$\l_n^\pm, n\in \Z$. There are a lot of results about inverse problems for
ZS-operators, see \cite{GK14}, \cite{K01}, \cite{K05}, and references therein.

\subsection{Preliminary results}
In order to study different invariants of the MS
we consider Manakov  operators $\cL$ on the real line given by
\[
\lb{do1} \tes
 \cL=iJ{d\/dx}+V, \qqq J=\ma 1&\!0\!\!\\ 0& -\1_{2}\am,\qqq  V=\ma 0& v^*\\
v&0\am=V^*,
\]
where $\1_d, d\ge 1$ is the identity $d\ts d$ matrix.
We assume that the 1-periodic $1\ts 2$ matrix
(vector) $v$ satisfies
\[
\lb{dv}
\begin{aligned}
v=(v_1,v_2)^\top\in \mH=L^2(\T)\os L^2(\T), \qqq \T=\R/\Z, \\
\end{aligned}
\]
where  $\mH$ is the complex Hilbert space equipped with the norm
$\|v\|^2=\!\!\int_0^1 (|v_1|^2+|v_2|^2)dx$.
Introduce the $3\ts 3$-matrix valued solutions $y(x,\l)$ of the Manakov system
\[
\lb{1}
\tes iJ y'+V y=\l y,\qqq  y(0,\l)=\1_{3}\qq \forall \ \l\in\C.
\]
Define the monodromy matrix $\p(\l)$ and its determinant $D(\t,\l)$ by
\[
\p(\l)=y(1,\l),\qqq \qq
D(\t,\l)=\det (\p(\l)-\t \1_{3}),\qqq \t,\l\in \C,
\]
which are entire functions of $\l$. An eigenvalue $\t$ of $\p(\l)$ is called a {\it
multiplier}, it is a root of the algebraic equation $D(\t,\l)=0$, where $\t,\l\in\C$. Let $\t_j=\t_j(\l), j\in \N_3=\{1,2,3\}$ be multipliers of
$\p(\l)$. It is known (see e.g., \cite{K10}) that the three
roots $\t_1(\cdot), \t_2(\cdot)$ and $\t_3(\cdot)$ constitute
three distinct branches of some analytic function $\t(\cdot)$
on the Riemann surface $\cR$.

The Manakov  operators $\cL=iJ{d\/dx}+V$ is self-adjoint and its spectrum $\s(\cL)$
is absolutely continuous (see \cite{G50})  and satisfies
\[
\lb{dsp} \s(\cL)=\s_{ac}(\cL)=\Big\{\l\in\R:|\t_j(\l)|=1,\
\text{some}\ j\in \N_3\Big\}=\gS_1\cup \gS_2\cup \gS_3,
\]
where  $\gS_j, j\in \N_3$ is the part of the spectrum of $\cL$
having the multiplicity $j$:
$$
\gS_3=\Big\{\l\in\R:|\t_j(\l)|=1,\ \forall\  j\in \N_3\Big\},\
\gS_1=\Big\{\l\in\R:|\t_j(\l)|=1,\ \text{only \ one }\ j\in
\N_3\Big\},
$$
and $\gS_2$ is the spectrum of $\cL$ with the multiplicity $2$.
In order to describe spectral problems it is convenient to define
the Lyapunov functions $\D_j(\l)$ on the Riemann surface $\cR$ by
\[
\lb{deL} \tes \D_j(\l)={1\/2}(\t_j(\l)+\t_j^{-1}(\l)), \qqq j\in
\N_3=\{1,2,3\}.
\]
In order to discuss periodic and anti-periodic eigenvalues we define
the corresponding determinants
$
D_\pm (\l)=D(\pm 1,\l)$.
The zeros of $D_{+}$ (or $D_{-}$) are the eigenvalues of  periodic (or
anti-periodic) problem for the equation $iJy'+Vy=\l y$. Denote by
$z_{2n,j}, (n,j)\in \Z\ts \N_3$ (and $z_{2n+1,j}$) the sequence of
zeros of $D_+$ (and $D_{-}$) counted with multiplicity such that
$$
z_{n,1}\le z_{n,2}\le z_{n,3}\le z_{n+1,1}\le z_{n+1,2}\le
z_{n+1,3}\le\dots .
$$
Note that $z_{n,j}$ is an eigenvalue of 2-periodic problem for the
equation $iJy'+Vy=\l y$.

In order to discuss the basic properties of the monodromy matrix we
consider the case $v=0$. The monodromy matrix $\p^o$, its multipliers
$\t_j^o$, the trace $T^o=\Tr y^o(1,\l)$, the Lyapunov functions $\D_j^o, j\in\N_3$ are given by
\[
\p^o(\l)=e^{-i\l J},\ \  \t_1^o=\t_2^o=e^{i\l },\ \
\t_3^o=e^{-i\l },\ \ T^o=e^{-i\l }+2e^{i\l },\qq \D_j^o=\cos \l.
\]
The corresponding determinants $D_\pm^o$ and its zeros (i.e.,  2-periodic
eigenvalues) $z_{n,j}^o$ have the form
$$
\tes  D_\pm^o(\l)=(e^{-i\l}\mp 1)(e^{-i\l}\mp 1)^2,\qq z_{n,j}^o=\pi n,
(n,j)\in \Z\ts \N_3.
$$
Below we need the following well-known results of Lyapunov (see p.109 [YS], \cite{K10}).

{\bf Theorem (Lyapunov )} {\it Let $v\in \mH$. Then
\\
\no i) If $\t=\t(\l)$ is a multiplier of $\p(\l)$  for
some $\l\in \C$, then $1/\ol \t(\l)$ is a multiplier of $\p(\ol \l)$.
In particular, if $\l\in \R$, then ${1/\ol\t(\l)}$ is the multiplier of
$\p(\l)$.
\\
ii) If $\t(\l)$ is a simple multiplier and $|\t(\l)|=1$ for some
$\l\in\R$, then $\t'(\l)\ne 0$.
\\
iii) The Riemann surface $\cR$ is $3$ or 2-sheeted. There
exists the Lyapunov function $\D$ on the Riemann surface $\cR$, where
each branch of the function $\D$ is given by
$\D_j={1\/2}(\t_j+\t_j^{-1})$, $j\in \N_3$.
\\
\no iv) Let $\D$ be real analytic on some interval $I=(a,b)\ss\R\cap
\cR$ and $\D(I)\ss (-1,1)$. Then $\D'(\z)\ne 0$ for each $\l\in I$
(the monotonicity property).

\no v) Each gap $(\l^-,\l^+)$ in the spectrum $\gS_3$ is an
interval $\ss \gS_1$ and its edges are periodic (anti-periodic)
eigenvalues or branch points of $\D$}.

We define the traces $T$ and $\wt T$ by
$$
T(\l)=\Tr \p(\l),\qqq \wt T(\l)=\ol T(\ol \l), \qq \l\in \C.
$$
Note that $T=\t_1+\t_2+\t_3$. We discuss the multipliers. Below we always assume that  $v\in \mH$.

\begin{theorem} \lb{T1}
i) Let $\t_j=\t_j(\l), \l\in \C, j\in \N_3$ be the multipliers of
$\p(\l)$. Then we have
\[
\lb{asm}
  \t_j(\l)=\t_j^o(\l)(1+o(1))\qq \as \qq |\l|\to \iy, 
 \qq \forall \  j\in \N_3,
\]
where $|\l-\pi n|\ge \d, $ for all $n\in \Z$ and for some $\d\in (0,1/2)$, and 
\[
\lb{5}
\begin{aligned}
  D(\t,\cdot)=-\t^3+\t^2T-\t e^{i\l } \wt
T+e^{i\l},\qq \t\in \C,
\end{aligned}
\]
\[
\lb{4}
\begin{aligned}
& \det \p=e^{i\l}=\t_1\t_2\t_3,
  \\
\tes
&
 \wt
T={1\/\t_1}+{1\/\t_2}+{1\/\t_3}=e^{-i\l}(\t_1\t_2+\t_2\t_3+\t_3\t_1).
\end{aligned}
\]
ii) Let $\gS_1, \gS_2, \gS_3$ be defined by \er{dsp}. Then the
spectrum of $\cL$ is given by
\[
\s(\cL)=\R=\gS_1\cup \gS_3,\qqq\qqq  \gS_2=\es.
\]

\end{theorem}

The coefficients of the polynomial $D(\t,\l)$ in $\t\in \C$ are
entire functions of $\l$. Due to \er{5} there exist exactly three
roots $\t_1(\cdot), \t_2(\cdot)$ and $\t_3(\cdot)$, which constitute
three distinct branches of some analytic function $\t(\cdot)$ that
has only algebraic singularities in $\C$, see, e.g.,  Ch.~8 in
\cite{Fo91}. Denote the set of all j-sheeted Riemann surfaces by $\mR_j, j\in \N$. Thus, in general, the function $\t(\cdot)$ is analytic
on some j-sheeted Riemann surface $\cR\in \mR_j, j\le 3$ and it has only a finite number of algebraic singularities any bounded domain.
In order to describe these points we introduce the {\it
discriminant} $\gD(\l),\l\in\C$, of the polynomial $D(\cdot,\l)$ by
\[
\lb{rd} \gD=-{e^{-2i\l}\/64}(\t_1-\t_2)^2(\t_1-\t_3)^2(\t_2-\t_3)^2.
\]
In fact, we have introduced the modified discriminant, since due to
the additional factor ${e^{-2i\l}}$ our function $\gD$ is real on the
real line. A zero of $\gD$ is a branch point of the corresponding
Riemann surface $\cR$. For us real branch points are  important.
We describe properties of the discriminant $\gD$. Note that the discriminant $\gD=0$ at $v=0$.

\begin{theorem}
\lb{Trho}
i)  The function $\gD(\l)$ is entire,
 real on the real line  and satisfies
\[
\lb{r} \gD=-{1\/64}\rt(T^2\wt T^2-4e^{-i\l }T^3-4e^{i\l } \wt T^3+18T\wt T-27\rt)
\qqq \forall \ \l\in \C,
\]
\[
\lb{esT}
 |T(\l)|\le 4e^{|\Im \l|}\ch \|v\|,
\]
\[
\lb{sro} \gS_1= \{\l\in\R: \gD(\l)<0\}, \qqq \gS_3= \{\l\in\R:
\gD(\l)\in [0,1]\}.
\]
ii) The function $\gD=0$ for some $v\in \mH$ iff we have $v=0$.

\end{theorem}

\no {\bf Remark.} i) This theorem describes the spectrum of $\cL$ in
terms of the function $\gD$. But it is difficult to study the
properties of $\gD$, since $\gD(\l)=0$ for all $\l$ at $v=0$.

\no ii)  Due to  Theorems \ref{T1}, \ref{Trho} we define gaps in the
spectrum $\gS_3$ of the multiplicity 3 by
\[
\lb{sm3} \gS_3=\R\sm \cup\g_n,\ \ \g_n=(\l_n^-,\l_n^+)\ne \es,\qq
... <\l_0^-\le \l_0^+<\l_1^-\le \l_1^+<\l_2^-\le \l_2^+<....
\]
Here $\g_n=(\l_n^-,\l_n^+)$ is a gap in the spectrum of $\gS_3$
It is important that
$\l_n^\pm$ are branch points for Lyapunov functions or 2-periodic
eigenvalues. We do not know results about  finite gap potentials.
We have no information about the number of gaps or asymptotics of gaps.

\begin{figure}
\tiny
\unitlength 1mm % = 2.845pt
\linethickness{0.4pt}
\ifx\plotpoint\undefined\newsavebox{\plotpoint}\fi % GNUPLOT compatibility
\begin{picture}(116.85,105.5)(0,0)
%lower graph
%coordinate lines
\put(6.25,71.9){\line(0,-1){58.4}}
\put(4.65,22.9){\line(1,0){107.6}}
\put(5.05,42.5){\line(1,0){105.4}}
\put(4.65,62.5){\line(1,0){105.4}}
\put(111.05,38.9){\makebox(0,0)[cc]{$\l$}}
\put(10.25,68.3){\makebox(0,0)[cc]{$\D(\l)$}}
\put(2.85,39.5){\makebox(0,0)[cc]{$0$}}
\put(3.05,19.1){\makebox(0,0)[cc]{$-1$}}
\put(2.85,59.3){\makebox(0,0)[cc]{$1$}}
%the curve
\qbezier(9.25,24.9)(20.65,20.2)(26.45,27.9)
\qbezier(20.85,42.5)(30.45,33.5)(26.45,27.9)
\qbezier(23.45,57.5)(15.15,49)(20.85,42.5)
\qbezier(38.25,59.9)(32.05,65.9)(23.45,57.5)
\qbezier(38.25,59.9)(42.85,56.1)(37.85,28.3)
\qbezier(37.85,28.3)(36.05,19)(31.05,26.9)
\qbezier(31.05,26.9)(29.65,28.3)(34.65,29.9)
\qbezier(34.65,29.9)(54.35,37.9)(57.65,43)
\qbezier(57.65,43)(71.35,62.4)(57.45,62.3)
\qbezier(57.65,43)(45.75,62.4)(57.45,62.3)
\qbezier(80.65,29.9)(60.95,37.9)(57.65,43)
\qbezier(84.25,26.7)(85.65,28.3)(80.65,29.9)
\qbezier(77.45,28.3)(79.25,19)(84.25,26.7)
\qbezier(77.05,60.1)(72.45,56.1)(77.45,28.3)
\qbezier(77.05,60.1)(82.85,65.9)(91.45,57.5)
\qbezier(91.45,57.5)(100.75,49)(94.05,42.5)
\qbezier(94.05,42.5)(84.45,33.5)(89.95,28.1)
\qbezier(107.05,24.9)(97.65,20.2)(89.95,28.1)
%upper graph
%coordinate lines
\put(5.05,87.5){\line(1,0){105.4}}
\put(109,83){\makebox(0,0)[cc]{$\l$}}
\put(8.25,83.5){\makebox(0,0)[cc]{$0$}}
\put(10.25,102.5){\makebox(0,0)[cc]{$\gD(\l)$}}
\put(20.65,90.25){\makebox(0,0)[cc]{$\l_0^+$}}
\put(27,90.25){\makebox(0,0)[cc]{$\l_{1}^-$}}
\put(34,90.25){\makebox(0,0)[cc]{$\l_{1}^+$}}
\put(39.5,90.25){\makebox(0,0)[cc]{$\l_{2}^-$}}
\put(53,90.25){\makebox(0,0)[cc]{$\l_{2}^+$}}
\put(97,90.25){\makebox(0,0)[cc]{$\l_5^-$}}
\put(90,90.25){\makebox(0,0)[cc]{$\l_{4}^-$}}
\put(84,90.25){\makebox(0,0)[cc]{$\l_{4}^+$}}
\put(76,90.25){\makebox(0,0)[cc]{$\l_{3}^-$}}
\put(64,90.25){\makebox(0,0)[cc]{$\l_{3}^+$}}
%the curve
\qbezier(18.5,87.5)(20.75,84.5)(28,87.5)
\qbezier(28,87.5)(30.5,88.75)(31,87.5)
\qbezier(31,87.5)(36.875,83.5)(40.75,87.5)
\qbezier(40.75,87.5)(47.5,95.375)(51.25,87.5)
\qbezier(51.25,87.5)(55.625,76.375)(58,75.75)
\qbezier(64.75,87.5)(60.375,76.375)(58,75.75)
\qbezier(75.25,87.75)(68.5,95.375)(64.75,87.5)
\qbezier(84.5,87.5)(79.125,83.5)(75.25,87.5)
\qbezier(88,87.5)(85.5,88.75)(84.5,87.5)
\qbezier(97.5,87.5)(95.25,84.5)(88,87.5)
\qbezier(18.5,87.5)(13.625,96.125)(4.25,98.25)
\qbezier(97.5,87.5)(102.375,96.125)(111.75,98.25)
%the spectrum
%coordinate line
\thinlines \put(6.25,79){\line(0,1){26.5}}
\put(4.25,3.5){\line(1,0){112.6}}
\put(100,5.75){\makebox(0,0)[cc]{$\gS_3$}}
%link lines
\thinlines
%\dottedline(18.5,87.5)(18.75,-16)
\multiput(18.297,86.797)(.002404,-.995192){85}{{\rule{.4pt}{.4pt}}}
%\end
%\dottedline(97.25,87.25)(97,-16.25)
\multiput(97.047,86.547)(-.002404,-.995192){85}{{\rule{.4pt}{.4pt}}}
%\end
%\dottedline(27.75,87.75)(27.75,-15.5)
\multiput(27.547,87.047)(0,-.992788){85}{{\rule{.4pt}{.4pt}}}
%\end
%\dottedline(88,87.5)(88,-15.75)
\multiput(87.797,86.797)(0,-.992788){85}{{\rule{.4pt}{.4pt}}}
%\end
%\dottedline(31.25,87.25)(31.25,-15.75)
\multiput(31.047,86.547)(0,-.990385){85}{{\rule{.4pt}{.4pt}}}
%\end
%\dottedline(84.5,87)(84.5,-16)
\multiput(84.297,86.297)(0,-.990385){85}{{\rule{.4pt}{.4pt}}}
%\end
%\dottedline(40.75,87.75)(40.75,-15.5)
\multiput(40.547,87.047)(0,-.992788){85}{{\rule{.4pt}{.4pt}}}
%\end
%\dottedline(75,87.5)(75,-15.75)
\multiput(74.797,86.797)(0,-.992788){85}{{\rule{.4pt}{.4pt}}}
%\end
%\dottedline(51.25,87.5)(51.75,-16)
\multiput(51.047,86.797)(.004808,-.995192){85}{{\rule{.4pt}{.4pt}}}
%\end
%\dottedline(64.5,87.25)(64,-16.25)
\multiput(64.297,86.547)(-.004808,-.995192){85}{{\rule{.4pt}{.4pt}}}
%\end
%spectral bands
\linethickness{4pt} \put(18.65,3.9){\line(1,0){9.2}}
\put(97.05,3.9){\line(-1,0){9.2}} \put(31.25,3.9){\line(1,0){9.8}}
\put(84.45,3.9){\line(-1,0){9.8}} \put(51.45,3.9){\line(1,0){8.2}}
\put(64.25,3.9){\line(-1,0){8.2}}
\end{picture}
\lb{fig1} \caption{\footnotesize The function $\gD$, the Lyapunov
function $\D$,  the spectrum $\gS_3$ and gaps $(\l_n^-,\l_n^+)$}
\end{figure}
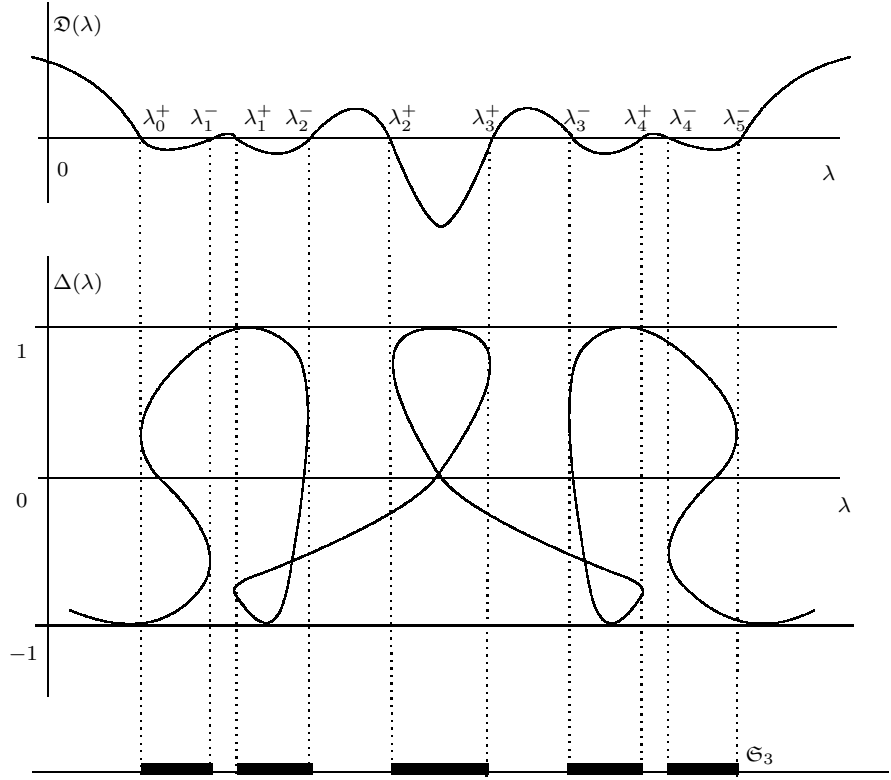

\subsection{Main results}
We discuss the properties of the Lyapunov functions and the Riemann surfaces.
 Let below  $\z\in \cR$ and let
$\dP(\z)=\l$ be the natural projection $\dP:\cR\to \C$.
The matrix $\mV=\int_0^1V ^2dx$ has eigenvalues $\gb_j\ge 0, j\in
\N_3$ given by:
\[
\lb{V23xx}
\begin{aligned}
\gb_3=\|v\|^2,\qq \gb_2={\|v\|^2+ \sqrt{\gb_o}\/2},\qq
\gb_1={\|v\|^2- \sqrt{\gb_o}\/2},
\qq
 \gb_o=(\|v_1\|^2-\|v_2\|^2)^2+4|c_{12}|^2,
\end{aligned}
\]
where $c_{12}=\int_0^1 v_1\ol v_2dx$. We define the function
$
\vr_\D=(\D_1-\D_2)^2(\D_1-\D_3)^2(\D_2-\D_3)^2.
$

\begin{theorem} \lb{T2x}
i)  Each function $\D_j, \t_j, j\in \N_3$ has the following asymptotics
as $|\l|\to \iy$:
\[
\begin{aligned}
\lb{LLa1} \D_j(\l)=\cos \l +o(e^{|\Im \l|}).
\end{aligned}
\]
If, in addition, $\Im \l\ge r|\Re \l|$ for any fixed $r>0$, then $\D_j, \t_j, j\in \N_3$ and $\gD$ have asymptotics
\[
\begin{aligned}
\lb{Djas}\tes
 \D_j(\l)=(1-{\gb_j+o(1)\/2i\l })\cos \l,
\end{aligned}
\]
\[
\begin{aligned}
\lb{ras1} \t_s(\l)=e^{i(\l-{\gb_s+o(1)\/2\l })}, \qqq s=1,2,\qq
\t_3(\l)=e^{-i(\l-{\gb_3+o(1)\/2\l})},
\end{aligned}
\]
\[
\lb{ras2} \tes \gD(\l)={e^{-i4\l}\/4^4\l^2}(\gb_o+o(1)).
\]
 ii)  If $\gb_1>0$ (given by \er{V23xx}), then the Riemann surface $\cR$ is 3-sheeted.
\\
\no iii) Let $\gf=\cT_--\sin \l$, where  $\cT_-={1\/2i}(T-\wt T)$. Then
the  functions $\gf, \vr_\D$ are entire, real on the real line, and satisfy
\[
\lb{dd1}
\begin{aligned}
D(e^{i\l},\l)=2ie^{i2\l}\gf(\l),\qq \forall \ \l\in \C,
\end{aligned}
\]
\[
\lb{dd2}
\begin{aligned}
\tes  4 \gD \ \gf^2=\vr_\D.
\end{aligned}
\]
If the Riemann surface is 3-sheeted, then  each zero of $\gf$ belongs to
$\gS_3$ and we have
\[
\lb{dd3} \gS_1= \{\l\in\R: \vr_\D(\l)<0\}, \qqq \gS_3= \{\l\in\R:
\vr_\D(\l)\ge 0\}.
\]

\end{theorem}

\no This theorem gives sufficient conditions, when
the Riemann surface  is the 3-sheeted.

Introduce the spaces $\ell^p, p\geq 1$ of sequences $f=(f_n)_{\Z}$
equipped with the norm $\|f\|_{p}^p=\sum |f_n|^p$. Below we write
$a_n=b_n+\ell^p(n)$ iff the sequence $(a_n-b_n)_{n\in \Z}\in
\ell^p(\Z)$.

Define a set $E_+$ of all periodic $z_{2n,j}$ and a set $E_-$ of all
anti-periodic eigenvalues $z_{2n+1,j}$ by
$$
E_+=\{z=z_{2n,j}, (n,j)\in \Z\ts \N_3\},\qqq E_-=\{z=z_{2n+1,j},
(n,j)\in \Z\ts \N_3\}.
$$

\begin{theorem}   \lb{T3}
The eigenvalues  $z_{n,j}, (n,j)\in \Z\ts \N_3$  have asymptotics as
$n\to\pm\iy$:
 \[
 \lb{a8}
 \begin{aligned}
& z_{n,j}=\pi n+\z_{n,j}|\wh v(\pi n)|+\ell^\d(n),\qqq  {\rm where }\qq
\z_{n,2}=0, \qq \z_{n,1}=1=-\z_{n,3},
\end{aligned}
\]
where  $\d>1$ and $ \wh v(\l)=\int_0^1 e^{i2\l x}v(x)dx$.
Moreover, if $D_\pm(0)\ne 0$, then the
functions $D_\pm$  have the Hadamard factorizations
\[
\lb{Dpm1}
\begin{aligned}
\tes D_\pm(\l)=D_\pm(0)e^{i{\l\/2}} \lim_{r\to +\iy}\prod_{|k_n|\leq r,  k_n\in
E_\pm}\Big(1-{\l\/k_n}\Big),
\end{aligned}
\]
where the constants  $D_+(0)=i2 \Im T(0),\  D_-(0)=2(1+ \Re T(0))$
and the product converges uniformly in every bounded disc. If the
spectrum $E_\pm$ and the constants $D_\pm(0)$ are given,  then
we can recover the discriminant $\gD$, the Riemann surface and the functions $T, \wt T$ by
\[
\lb{fjd1}
\begin{aligned}
\tes T={D_-+D_+\/2}-e^{i\l},\qqq
  \wt T=e^{-i\l}{D_--D_+\/2}-e^{-i\l}.
\end{aligned}
\]

\end{theorem}

{\bf Reamrk.} If we know the periodic spectrum for  ZS-operators, then we can recover the anti-periodic one. This defines the corresponding Riemann surface. For the Manakov system it is not true. In order to recover the Riemann surface we need periodic and anti-periodic eigevalues.

Define the set $
\mH_o=\Big\{v\in \mH: v=ue,\qq (u,e)\in L^2(\T)\ts \C^2, \qq |e|=1\Big\}
$.
Here if $v=(v_1,v_2)^\top\in \mH_o$, then $v_1=0$ or $v_2=0$ or $v_1=\a v_2$  for some $ \a\in \R$.

 \begin{proposition}
\lb{Tpr} Let $v\in \mH_o$, i.e., $v=ue$ for some constant vector $e\in \C^2, |e|=1$  and $u\in L^2(\T)$. Then the Manakov operator $\cL$ is unitary equivalent to the operator $\cL_u=\cL_{zs}\os {d\/idx}$, where $\cL_{zs}$ is the Zakharov-Shabat operator on $L^2(\R,\C^2)$ given by \er{ddzs}, and $\cL, \cL_u$ satisfy
\[
\lb{uzs1}
\begin{aligned}
& \tes \cL_u =\cU \cL \cU^*=iJ{d\/dx}+V_u, \qqq V_u=V_{zs}\os 0, \qq \cU=1\os U,
\\
& U^*=( e, e_o),\qq e_o\in \C^2,\qq |e_o|=1, \qq e_o \perp e.
\end{aligned}
\]
Furhtermore, the monodromy matrices for $\cL_u$ and $\cL$ are unitary equivalent, their multipliers, periodic eigenvalues are the same,
the Riemann surface $\cR$ is 2-sheeted and $\gS_3(\cL)=\s(\cL_{zs})$.
\end{proposition}

 If $v\in \mH_o$, then the Riemann surface $\cR$ degenerates into two components $\cR_{zs}$  and the complex plane, where $\cR_{zs}$ is
the Riemann surface for ZS-operators. Thus the Riemann surface $\cR$ is 2-sheeted. We describe all 2-sheeted Riemann surface and their
branch points.

 \begin{theorem}
\lb{T2sR}
Let an interval $\o\ss \g_o$
for some open gap $\g_o=(\l^-, \l^+)$ in the spectrum
$\gS_3$.
 Then the  following relations concerning the Riemann surface $\cR$ hold true:
\[
\lb{uzs2}
\begin{aligned}
v\in \mH_o \qq  \Leftrightarrow
\qq \cR\in \mR_2\qq \Leftrightarrow \qq  \vr_\D=0 \qq  \Leftrightarrow\qq \gf=0 \qq \Leftrightarrow \D_3(\o)\ss \R \sm[-1,1]
\\
\Leftrightarrow\qq \D_2(\o)\ss \R \sm[-1,1] \qq \Leftrightarrow \qq  \D_2=\D_3 \qq \Leftrightarrow \qq
\t_2\t_3=1 \qq \Leftrightarrow \qq  \t_1=e^{i\l},
\end{aligned}
\]
or
\[
\lb{uzs3}
\begin{aligned}
 v\in \mH_o \qq  \Leftrightarrow
 \qq \cR\in \mR_2\qq \Leftrightarrow \qq  \vr_\D=0 \qq  \Leftrightarrow\qq \gf=0\qq \Leftrightarrow \D_3(\o)\ss \R \sm[-1,1]
\\
\Leftrightarrow\qq  \D_1(\o)\ss \R \sm[-1,1] \qq \Leftrightarrow \qq  \D_1=\D_3 \qq \Leftrightarrow \qq
\t_1\t_3=1 \qq \Leftrightarrow \   \t_2=e^{i\l}.
\end{aligned}
\]

\end{theorem}
\no {\bf Remark.} 1) The two-sheeted Riemann surface of $\cL$ is the Riemann surface
for ZS-operators. Thus all their branch points are the 2-periodic eigenvalues
for ZS-operators.

\no 2) From this theorem we deduce that
$\D_j(\g_n)\ss \R\sm (-1,1)$ for some $(j,n)\in \N_3\ts \Z$ iff the Riemann surface is 2-sheeted. Thus for each open gap $\g_n$ there exists a branch $\D_j$
such that $\D_j(\g_n)\ss \R\sm (-1,1)$.

\no 3) If the Riemann surface $\cR$ is two-sheeted, then we can solve the inverse problem in terms of gap lengths, using the results of the author \cite{K05} for ZS-operators. Or via the Greber-Kappeler  results \cite{GK14} we can discuss the action-angle bijection.

 \begin{corollary}
\lb{T6}  The following relations hold true:
\[
\lb{61}
\begin{aligned}
v=0\qq \Leftrightarrow \qq  \gS_3=\R \qq  \Leftrightarrow
\qq \gD=0 \qq  \Leftrightarrow\qq \D_1=\D_2\qq  \Leftrightarrow\qq \t_1=\t_2.
\end{aligned}
\]
\end{corollary}

We discuss the MS for two sheeted Riemann surfaces.

\begin{corollary} \lb{T7}
i) Let $v$ be the solution to MS  $i\dot v=-v''+|v|^2v$,
under the initial condition $v_2=pv_1$  at $t=0$ for some constant $p\in \C\sm \{0\}$.
Then $p$ is a constant of motion,  $v_2=pv_1$ for any time $t>0$,  and the MS  transforms to the two separated NLS equations
\[
\lb{71}
i\dot y_j=-y_j''+ 2|y_j|^2y_j,\qq j=1,2, \qq y_1=(1+|p|^2)^{1\/2}v_1,\qq  y_2=e^{i\arg p}y_1.
\]
ii) Let $y_1,y_2$ be the solutions of the two separated NLS equations
$i\dot y_j=-y_j''+ 2|y_j|^2y_j,\qq j=1,2$ and $y_2=e^{i\vp}y_1$ for some
constant $\vp\in \R$. Then the vector-function $v=(v_1,v_2)$ given by
\[
\lb{72}
v_1=y_1/(1+r^2)^{1\/2}, \qq v_2=re^{i\vp}v_1,
\]
for some $r>0$ is the solutions of the MS.

\end{corollary}

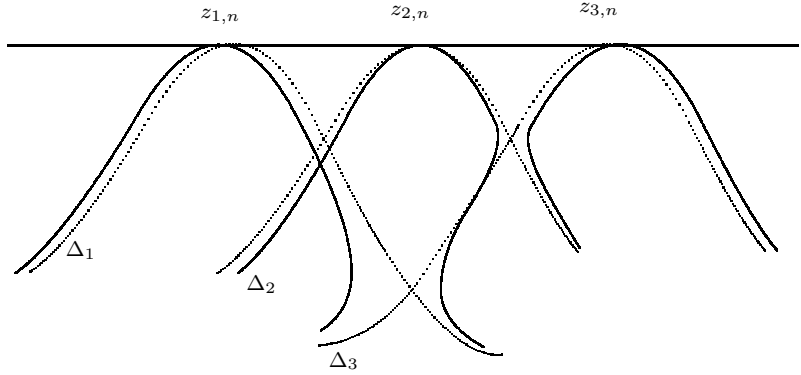
\begin{figure}
\tiny
\unitlength 0.8mm % = 2.845pt
\linethickness{0.4pt}
\ifx\plotpoint\undefined\newsavebox{\plotpoint}\fi % GNUPLOT compatibility
\begin{picture}(143,83.375)(0,0)
\put(3,70){\line(1,0){130}} \put(153,36){\line(0,1){.25}}
%%%%%%%%%%%%unperturbed lines
%%%%%%%%%%%first
\qbezier[50](25.5,56.75)(40.5,83.375)(53.5,57.5)
\qbezier[50](25.5,56.75)(14.125,37.625)(6.75,32.5)
\qbezier[45](53.5,57.5)(61.75,40.875)(65,36)
\qbezier[55](65,36)(76.375,17.875)(84.25,18.75)
%%%%%%%%%%%second
\qbezier[50](56.25,56.5)(44.875,37.375)(37.5,32.25)
\qbezier[50](56.25,56.5)(71.25,83.125)(84.25,57.25)
\qbezier[50](84.25,57.25)(92.5,40.625)(96.75,35.75)
%%%%%%%%%%%%%third
\qbezier[45](70.25,30.5)(63.25,21.125)(54.25,20.25)
\qbezier[45](87,56.75)(82.75,49.25)(70.25,30.5)
\qbezier[50](87,56.75)(102,83.375)(115,57.25)
\qbezier[50](115,57.25)(123.25,40.875)(127.5,36)
%%%%%%%%%%%perturbed lines
%%%%%%%%%%first
\qbezier(23.5,57)(11.625,37.375)(4.25,32.25)
\qbezier(23.5,57)(38,83.125)(51,57.25)
\qbezier(51,57.25)(60.125,39.875)(59.5,31.25)
\qbezier(59.5,31.25)(59,25.625)(54.5,22.75)
%%%%%%%%%%second
\qbezier(58.175,57)(47.487,37.375)(40.85,32.25)
\qbezier(58.175,57)(71.225,83.125)(83.25,57)
\qbezier(83.25,57)(85,53)(78.75,43)
\qbezier(78.75,43)(74,35.625)(74.25,28.75)
\qbezier(74.25,28.75)(74.75,23.625)(81.25,20)
%%%%%%%%%%third
\qbezier(89.5,50.25)(90.75,46.625)(97,36.5)
\qbezier(89,56.5)(87.875,54.5)(89.5,50.25)
\qbezier(89,56.5)(104,83.375)(117,57.5)
\qbezier(117,57.5)(125.25,40.875)(129.5,36)
\put(15,36){\makebox(0,0)[cc]{$\D_1$}}
\put(44.75,30.5){\makebox(0,0)[cc]{$\D_2$}}
\put(58.25,17.75){\makebox(0,0)[cc]{$\D_3$}}
\put(38,74.75){\makebox(0,0)[cc]{$z_{1,n}$}}
\put(69.25,75.25){\makebox(0,0)[cc]{$z_{2,n}$}}
\put(100.25,75.75){\makebox(0,0)[cc]{$z_{3,n}$}}
\end{picture}
\lb{fig4}
\caption{\footnotesize the Lyapunov functions $\D_j$ and
$z_{j,n}, j\in \N_3$ are the periodic eigenvalues}
\end{figure}

\subsection{History }
There is an enormous literature on the scalar Hill operator,
including the inverse spectral theory, see \cite{MO75}, \cite{GT87},
\cite{KK97}, \cite{K99}, and for the periodic distributions \cite{K03}, and see references therein. There are lot of results about periodic systems, Lyapunov,  Krein \cite{Kr55}, \cite{Kr57}, Gel'fand-Lidskii
\cite{GL55} and, in particular,
about first order systems \cite{K08}, \cite{K10},
 and second order systems  \cite{CK06},  \cite{CG02},
\cite{C00}, \cite{K10}.  We mention some papers relevant for our
context.  The matrix potentials pose new problems. The Riemann
surfaces are multi-sheeted and have complicated structures. The
branch points can be real and complex. The spectral bands can
have various multiplicity and their end points are
branch points or periodic eigenvalues. Their control is not
simple. Until now the inverse problem  is solved only for the Schr\"odinger operator with matrix-valued potential on the finite interval \cite{CK06q}, \cite{CK09} (including characterization) and for first order systems \cite{M99}, \cite{M15} (only the uniqueness). Unfortunately, there are no results about periodic case.

Schr\"odinger operators with $d\ts d$ matrix valued
periodic potentials are discussed in \cite{CK06}  and the case $d=2$ in \cite{BBK06}. Lyapunov functions are discussed and the asymptotics of branch
points and periodic eigenvalues are determined. The estimate of gap
lengths in terms of potentials is obtained. In order to prove this
estimate the conformal mapping (the averaged quasimomentum) constructed is constructed and the basic properties are determined. In \cite{K10} similar results were derived for the Lax type operators (for the matrix NLSE), where the matrix $J=\1_{m}\os -\1_{n}$ for some $m,n\in \N$. The case $m=n$ is
considered in \cite{K08}.  When we study the  first order
systems we have additional problems, associated with the integral
equation for fundamental solutions, where the kernel is a bounded function on the period. Recall that for Schr\"odinger operators the corresponding  kernel has asymptotics $O(1/\l)$ as $\l\to +\iy$. It helps a lot.
We shortly discuss results from
\cite{CK06}, \cite{K08}, \cite{K10}:

\no $\bu$ The Lyapunov functions and multipliers on the Riemann
surfaces are considered. The spectrum of operators in terms of these
functions is described.
\\
$\bu$ It was shown that the endpoints of the gaps are periodic or
antiperiodic eigenvalues or are branch points of the Lyapunov
function.
\\
$\bu$ The existence of real and complex branch points has been proven
for specific potentials.
\\
$\bu$ The asymptotics of eigenvalues of the periodic and
antiperiodic problems and of branch points of the Lyapunov function
are derived.
\\
$\bu$ The  average  quasimomentum as a conformal mapping was
constructed and its properties are described. It generalizes the
global quasimomentum for Hill operators \cite{Fi75}, \cite{MO75}.
Note that useful results about the conformal mappings were obtained
in \cite{KK05}. In fact,  some
spectral problem for the differential operator with periodic matrix
coefficients are reformulated as problems of conformal mapping theory.
\\
$\bu$  Trace formulas (similar to those in the scalar case for Hill
operators \cite{KK95}, \cite{K97}) are obtained.
\\
$\bu$ The spectral gap lengths are estimated in terms of the norm of
potentials.
\\
$\bu$ A similar analysis for periodic systems of difference
equations was performed by Korotyaev and Kutsenko \cite{KKu08},
\cite{KKu09}.
\\
\no
$\bu$ Finally we discuss  3-order
scalar operators, see e.g., \cite{M81}, \cite{BK12}, \cite{BK13}, and describe  their properties.
Such operators are used to integrate  the Boussinesq equation \cite{M81}.
Consider the self-adjoint third order operator with 1-periodic coefficients on
the real line \cite{BK12}, \cite{BK13}. Its spectrum is absolutely continuous, covers the real line
and is a union of bands of multiplicity three separated by gaps of multiplicity one.  The spectrum of multiplicity three is bounded.
The Lyapunov function is analytic on the 3-sheeted Riemann surface.
Its branch points are real or can be complex.
Note that in order to study the good Boussinesq equation we need to is considere another non-self adjoint third order operator. Various properties of such operators (inverse problems, divisors and so on) were discussed in \cite{M81}, \cite{BK24}.

\section { Monodromy matrices and examples \lb{Sec2}}
\setcounter{equation}{0}

\subsection{Example 1. Zakharov-Shabat operators } Consider scalar defocusing periodic NLS
equations \er{VSE1} with a scalar function $u\in L^2(\T)$. To study
this equation Zakharov and Shabat introduced the Z-S operator $\cL_{zs}$ on $L^2(\R,\C^2)$ given by
\[\lb{zs}
\cL_{zs}=iJ_{zs} {d\/ dx}+ V_{zs},\ \ \ \  J_{zs} =\ma 1&0\\ 0&-1\am, \
\ \ \ \ V_{zs}= \ma 0 &  \ol u\\  u & 0\am,\qq u\in L^2(\T).
\]
 Recall known facts about the Z-S operator (see, e.g.,
\cite{LS91}). The spectrum of $\cL_{zs}$ are purely absolutely
continuous and consists of spectral bands separated by the gaps:
\[
\lb{zs12}
\s(\cL_{zs})=\R\sm \cup \g_n, \qq \g_n=(\l_n^-,\l_n^+),\qq
\l_n^\pm=\pi n+o(1)\ \ \as \ n\to \pm \iy.
\]
Introduce the $2\ts 2$-matrix valued solutions $y_{zs}(x,\l)$ of
the Zakharov-Shabat system
\[
\lb{zs1} iJ_{zs}y_{zs}'+V_{zs}y_{zs}=\l y_{zs},\qqq
y_{zs}(0,\l)=\1_{2}.
\]
The $2\ts 2$ matrix valued solutions $y_{zs}(1,\l)$ is entire and  has two
multipliers $\t_{zs}, \t_{zs}^{-1}$, which are analytic on the
2-sheeted Riemann surface. The Lyapunov function has the form $
\D_{zs}(\l)={1\/2}(\t_{zs}(\l)+\t_{zs}^{-1}(\l))$, is entire and
satisfies $\D_{zs}(\l)={1\/2}\Tr y_{zs}(1,\l) $. It defines the
band-gap structure of the spectrum.
Here the end of the gap $\g_n=(\l_n^-,\l_n^+)$ are eigenvalues of
 2-periodic problem for the equation $iJ_{zs}y_{zs}'+V_{zs}y_{zs}=\l y_{zs}$.
 The eigenvalues  $\l_{2n}^\pm$ are eigenvalues of
 1-periodic problem for the equation $iJ_{zs}y_{zs}'+V_{zs}y_{zs}=\l y_{zs}$.
 And the eigenvalues  $\l_{2n+1}^\pm$ are eigenvalues of
 anti-periodic problem for the equation $iJ_{zs}y_{zs}'+V_{zs}y_{zs}=\l y_{zs}$. Moreover, we have
 $$
 \D_{zs}(\l_{n}^\pm)=(-1)^n,\qq n\in\Z.
 $$
If $\g_n=\es$, then $\l_{n}^\pm$ is a double  eigenvalues.
For each $u$ there exists a unique conformal mapping (the quasi-momentum) $k:\C\sm\cup \bar \g_n\to \cK $ such that (see \cite{Mi78}, \cite{KK95})
$$
\cos k(\l)= \D_{zs}(\l),\ \  \l\in \C\sm\cup \bar \g_n, \ \ \ \
\cK=\C\sm\cup \G_n ,\ \  \  \G_n=(\pi n-ih_n,\pi n+ih_n),
$$
$$
k(\l)=\l-{\|v\|^2+o(1)\/ 2\l}\ \ \as \ \ \ \ \l\to i\iy , \ \ {\rm
and }\ \ \ \ k(\l)=\l+o(1)\qq \as \ \ \l\to \pm\iy,
$$
 where $\G_n$ is the excised vertical
cut and the height $h_n\geq 0$ is defined by the equation $\ch h_n
=\max_{\l\in \g_n} |\D_{zs}(\l)| \geq 1$. First classical results about
such mappings  was obtained by Hilbert for a finite number of cuts
(see, e.g., \cite{J58}). For Hill operators such conformal mappings
were constructed  simultaneously by Firsova \cite{Fi75} and by Marchenko-Ostrovski \cite{MO75}. Various properties of such conformal mappings, including trace formulas were determined in \cite{KK95}, \cite{K96}, \cite{K97}, \cite{K00}, \cite{K05}, \cite{K06} for Hill and ZS-operators. For example, for ZS-operators the estimate of gap lengths in terms of potentials was obtained
\[
\lb{KZS} \tes
 {1\/ \sqrt 2}g\leq \|u\|\leq 2g\big(1+g\big),
\]
by EK \cite{K05}, where $g=(\sum |\g_n|^2)^{1/2}$. Moreover, the
estimates of Hamiltonians for NLS equetions in terms of action
variables were determined in \cite{K05}.

\subsection
{The vector  $v\in \mH_o$} Here we consider the second example with $v\in \mH_o$.

{\bf Proof of Proposition \ref{Tpr}.}
 Let $v=ue\in \mH_o$ for some constant vector $e\in \C^2, |e|=1$  and $u\in L^2(\T)$. We show that $\cL$ is unitary equivalent to the operator $\cL_u=\cL_{zs}\os {d\/idx}$, where $\cL_{zs}$ is the Zakharov-Shabat operator on $L^2(\R,\C^2)$ given by \er{ddzs}.
 Let $\cU=1 \os U \in \M^3 $, where the matrix $U\in \M^2$ is  unitary and given by
 $$
 U^*=( e, e_o),\qq e_o\in \C^2,\qq |e_o|=1, \qq e_o \perp e.
 $$
 Define the operator $\cL_u= \cU \cL \cU^*$ and the vector $w=U v=uU e=(u,0)^\top$. We have
 \[
\lb{uzs1x}
\begin{aligned}
\cL_u= U \cL U^*=i{d\/dx}J +V_u ,\qqq J= \cU J \cU^*=J,\qqq V_u= \cU V \cU^*.
\end{aligned}
\]
We compute the operator $V_u$:
$$
 V_u= \cU V \cU^*=\ma 1 & 0\\ 0 & U\am  \ma 0 & v^*\\ v &0\am   \ma 1 & 0\\ 0 & U^*\am =\ma 0 & v^*\\ Uv & 0      \am        \ma 1 & 0\\ 0 & U\am =
 \ma 0 & w^*\\ w & 0 \am =V_{zs}\os 0.
$$
Thus we have $\cL_u=\cL_{zs}\os {d\/idx}$ and $V_u=V_{zs}\os 0$.
The operators $\cL, \cL_u$ are unitary equivalent and their monodromy matrices are unitary equivalent also. Then their multipliers, periodic eigenvalues are the same. The multipliers and the Lyapunov functions have the form
\[
\lb{mtt}
\t_1=e^{i\l},\ \  \t_2=e^{ik}, \ \ \t_3=e^{-ik},
\ \ \D_1=\cos \l, \ \ \D_2=\D_3=\D_{zs}=\cos k(\l),
\]
 where $k=k(\l)$ is
the quasimomentum for the operator $\cL_{zs}$. Recall  that for
ZS-operators  the Lyapunov function $\D_{zc}=\cos k(\l)$.  We compute $T$ and $\gD$:
\[
T=e^{-ik}+e^{ik}+e^{i\l}=2\cos k+e^{i\l}=2\D_{zc}+e^{i\l}
\]
and
\[
\lb{r0}
\begin{aligned}
& \tes
64\gD=-e^{-i2\l}(e^{-ik}-e^{ik})^2(e^{ik}-e^{i\l})^2(e^{i\l}-e^{-ik})^2
= 64 \sin^2{k} \sin^2{k-\l\/2}\sin^2{k+\l\/2}
\\
&=16\sin^2{k}(\cos k-\cos
\l)^2=16(1-\D_{zs}^2(\l))(\D_{zs}(\l)-\cos \l )^2.
\end{aligned}
\]
 Then from \er{sro} we have  that $
\gS_3=\s(\cL_{zs})=\{\l\in \R: \D_{zs}^2(\l)<1\}. $ We also have
zeros of the factor $(\D_{zs}(\l)-\cos \l )^2\ge 0$ on $\R$ and it
does not create the gaps in the spectrum $\gS_3$. Similar arguments
give
$$
\gf=0,\qq \vr_\D=0,\qq D_\pm=2(1\mp\D_{zs})(e^{i\l}-1).
$$
 Then from the identities
$D(\t,\l)=-(\t-\t_1)(\t-\t_2)(\t-\t_3)$ and \er{mtt}  we have
\[
D(\t,\l)=-(\t-e^{i\l})(\t^2-2\t \D_{zs}+1).
\]
 The Lyapunov function $\D_1=\cos \l$, which implies $z_{n,1}=\pi n, n\in \Z$. The Lyapunov function $\D_2=\D_3=\D_{zs}$
and it gives the spectral bands of $\gS_3$ and the corresponding gaps
$\g_n=(\l_n^-,\l_n^+), n\in \Z$.
Here $\l_n^\pm$ are 2-periodic eigenvalues and they have asymptotics
\[
\lb{RS23}
\begin{aligned}
 z_{n,2}=\pi n, \qq  z_{n,2}=\l_n^-,\qq z_{n,3}=\l_n^+,
 \qq
 \l_n^\pm =\pi n\pm |\wh v(\pi n)|+\ell^\d(n)
\end{aligned}
\]
as $ n\to \pm  \iy$, for some $\d>1$ and if $g=(\sum
|\g_n|^2)^{1/2}$, then we have estimates
\[
\lb{RS24} \tes
 {1\/ \sqrt 2}g\leq \|v\|\leq 2 g\big(1+g\big).\qqq \qqq \qqq \qqq \BBox
\]

\subsection{Properties of matrices}
 We begin with some notational
convention. A vector $h=(h_n)_1^d\in \C^d$ has the Euclidean norm
$|h|^2=\sum_1^d|h_n|^2$, while a $d\ts d$ matrix $A$ has the
operator norm given by $|A|=\sup_{|h|=1} |Ah|$. Note that $|A|^2\le
\Tr A^*A$.
\\
$\bu$  We define the space $\C^{n,m}$ of all $n\ts m$ complex
matrices and let $\M^{d}=\C^{d,d}$.
\\
$\bu$ For any matrices $A, B\in \M^{d}$ the following identities
hold true:
\[
\lb{AB}
 \Tr AB=\Tr BA, \ \ \ \ \ \ \ \ \ol{\Tr A}=\Tr A^*.
\]
$\bu$ Suppose that a matrix-valued function $A(\l), \l\in\mD$ is
analytic in the domain $\mD \ss {\C}$ and has an inverse
$A^{-1}(\l)$ for any $\l\in \mD$. Then for any $\l\in \mD$ the
function $\det A(\l)$ satisfies
\[
\lb{Dd}
 {d\/d\l}\det A(\l)=\det A(\l){\rm Tr}\,A (\l)^{-1}A'(\l).
\]
$\bu$ If $A\in \M^d$, then the following identity holds true:
\[
\lb{Das}\tes  \det (I+A)=\exp \Big(\Tr A-\Tr {A^2\/2}+\Tr
{A^3\/3}+..\Big)\ \ \as \qq \|A\|<1.
\]
$\bu$ Let $\mA$ be the class of all matrices $A=\ma
0_1&a_1\\a_2&0_2\am$, where $a_1,a_2^* \in \C^{1,2}$ and $0_d=0\in
\M^d$. Below we need following  identities for any  $A,B, C\in\mA$
and $\l\in \C$:
\[
\lb{23} J A=-AJ, \ \ \ \ \ \ e^{\l J} A=Ae^{-\l J},
\]
\[
\lb{213} ABC,\ JA ,\ e^{\l J}A \in \mA,\ \ \ \ \
\]
\[
\lb{214} AB=\ma a_1b_2&0\\0&a_2b_1\am ,
\]
\[
\lb{215} \Tr A=0,\ \   \Tr JA^n=0,\ \  \forall \ n\in \N.
\]

\begin{lemma}  \lb{Tv}
i) The matrix $V_x, x\in \T$ has eigenvalues $z_\pm =\pm |v(x)|,
z_o=0, $ and $|V_x|=|v(x)|$.

\no ii)  Let $\mV=\int_0^1V_x^2dx$. Then  the following identities hold
true:
\[
\lb{V22xx}
\begin{aligned}
\mV=\int_0^1V_x^2dx
           =\ma   c & 0 & 0\\
          0 & c_1 & \ol c_{12} \\
           0 &  c_{12} & c_2 \am,
           \qq
           \Tr \mV=2\|v\|^2,
\end{aligned}
\]
\[
\lb{trv}
\begin{aligned}
 \Tr e^{-i\l J}J\mV=-i2(\sin \l)\|v\|^2.
\end{aligned}
\]
where $c_j=\int_0^1|v_j|^2dx,\qq j=1,2, c=c_1+c_2,\qq
c_{12}=\int_0^1 v_1\ol v_2dx$. Moreover, the matrix $\mV$ has
eigenvalues $\gb_j\ge 0, j\in \N_3$ given by:
\[
\lb{V23} \tes \gb_3=c,\qq \gb_2={1\/2}(c+ \sqrt{\gb_o}),\qq
\gb_1={1\/2}(c- \sqrt{\gb_o}),\ \  \gb_o= (c_1-c_2)^2+4|c_{12}|^2\ge
0.
\]

\end{lemma}

 \no {\bf  Proof.} The proof of i) is simple. ii) We have
$$
\mV=\int_0^1V ^2dx=\int_0^1 \ma |v|^2 & 0 & 0\\
          0 &  |v_1|^2 &   \ol v_1 v_2\\
           0 & v_1\ol v_2  & |v_2|^2 \am dx
           = \ma   c & 0 & 0\\
          0 & c_1 & \ol c_{12} \\
           0 &  c_{12} & c_2 \am,
           \qq
           \Tr \mV=2\|v\|^2.
$$
We show \er{trv}. We have $e^{-i\l J}J=(e_{-1}, -e_1\1_2)$ where
$e_{x}=e^{ix\l J}$ and then
$$
\Tr \ma   e_- & 0 & 0\\
          0 & -e & 0 \\
           0 & 0  & -e \am
           \ma   c & 0 & 0\\
          0 & c_1 & \ol c_{12} \\
           0 &  c_{12} & c_2 \am=e_{-1}c-e_1c_1-e_1c_2=-2i \|v\|^2\sin \l,
$$
which yields \er{trv}. From \er{V22xx} we obtain \er{V23}.
 \BBox

\subsection{Fundamental solutions.}

We begin to study the fundamental $3\ts 3$-matrix solutions
$y(x,\l)$ of the equation
\[
\lb{DE1} iJy'+V y=\l y,\qqq  y(0,\l)=\1_{3}\qq \forall \ \l\in\C.
\]
This solution $y$ satisfies the standard integral equation
\[
\lb{26} y(x,\l)=y_0(x,\l)+i\int_0^x e^{i\l (s-x)J}JV_sy
(s,\l)ds,\ \ \ y_0(x,\l)=e^{-i\l xJ}.
\]
It is known that equation  \er{26}  has a solution as a power series in
$V$ given by
\[
\lb{27} y(x,\l)=\sum_{n\ge0}y_n(x,\l), \ \ \
y_n(x,\l)=i\int_0^xe^{i\l (s-x)J}JV_sy_{n-1}(s,\l)ds,\ \ n\ge1.
\]
 Using \er{27}, \er{23} we have for the first term:
\[
\lb{28}
\begin{aligned}
 y_1(x,\l)=i\int_0^xe^{i\l (s-x)J}JV_se^{-i\l sJ}ds=
i\int_0^xe^{i\l (2s-x)J}JV_sds,
\end{aligned}
\]
and the second term:
\[
\lb{29} y_2(x,\l)=
\int_0^xds_1\int _0^{s_1}e^{i\l J\ga_2}V_{s_1}V_{s_2}ds_2,
\qqq
V_{s_1} V_{s_{2}}=\ma  v^*(s_1)v(s_2)& 0\\
 0 & v(s_1)v^*(s_2)\am,
\]
where $\ga_2=2(s_1-s_2)-x$. Proceeding by induction and setting
$\ga_n$ we obtain
\[
\lb{210}
\begin{aligned}
&  y_{n}(x,\l)=i^n\int_{G_n(x)} e^{i\l J\ga_n(x,s)}JV_{s_1}\dots JV
_{s_n}ds,\qq s=(s_1,..,s_n)\in \R^n,
\\
&  G_n(x)=\{0< s_n<... s_2<s_1< x\}\ss \R^n,\ \
\ga_n(x,s)=2(s_1-s_2+\dots +(-1)^ns_n)-x,
\end{aligned}
\]
and, in particular,
\[
\lb{211} y_{2n}(x,\l)=\int_{G_{2n}(x)} e^{i\l
J\ga_{2n}(x,s)}V_{s_1}\dots V_{s_{2n}}ds,
\]
\[
\lb{212} y_{2n+1}(x,\l)=i J\int_{G_{2n+1}(x)} e^{-i\l
J\ga_{2n+1}(x,s)}V_{s_1}\dots V_{s_{2n+1}}ds,
\]
 where
\[
\lb{212x} V_{s_1}\dots V_{s_{2n}}=\ma v^*(s_1)v(s_2)\dots
v(s_{2n}) &0\\0& v(s_1)v^*(s_2)\dots v^*(s_{2n})\am .
\]
 Now we show  the basic properties of the function $y$.

\no  \begin{lemma}  \lb{Tfs} For each $(\l,v)\in \C\ts \mH$ there
exists a unique solution $y(x,\l)$ of the equation
\er{26} given by \er{27} and series \er{27} converge uniformly on
bounded subsets of $\R\ts\C\ts \mH$. For each $x\in [0,1]$ the
function $\p (x,\l)$ is entire on $\C$ and  satisfies:
\[
\lb{217}
\begin{aligned}
 |y_n(x,\l)|\le e^{x|\Im \l|}{\|v\|^n\/n!},
 \qqq
 |y(x,\l)|\le e^{x|\Im \l|+\|v\|},
 \end{aligned}
\]
 for any
$(n,x,\l)\in \N\ts [0,1]\ts \C$,
where $\|v\|^2=\int_0^1(|v_1(x)|^2+|v_2(x)|^2)dx$ and
\[
\lb{218} |y(x,\l)-\sum_0^{n-1}y_j(x,\l)|\le {\|v\|^n\/n!}e^{x|\Im
\l|+\|v\|},
\]
\[
\lb{219} y(x,\l)-e^{-i\l xJ}=o(1)e^{x|\Im \l|}\ \qqq as \qq |\l|\to
\iy,
\]
uniformly on bounded subsets of $\R\ts \mH$. Moreover, if the
sequence $v^{(m)}\to v$ weakly in $\mH$, as $n\to \iy$, then
$y(x,\l,v^{(m)})\to y(x,\l,v)$ uniformly on bounded subsets of
$[0,1]\ts \C$.
\\
Furthermore, let sequence $ |\l_n-\pi n|\leq
{\pi\/4}, n\in \Z$ and $\l_n\to \pm \iy$ as $n\to \pm \iy$. Then  for any fixed $\d>2$ the following asymptotics
\[
\lb{aspzn}
\begin{aligned} y(x,\l_n)=y_o(x,\l_n)+y_1(x,\l_n)+y_{(2)}(x,\l_n),
\\
|y_1(x,\l_n)|=\ell^{\d}(n) ,\qq |y_{(2)}(x,\l_n)|=\ell^{\d/2}(n),
\end{aligned}
\]
as $n\to \iy $ hold, uniformly on $[0,1]\times\{|\l_n-\pi n|\leq
{\pi\/4}\}\ts \{\|v\|\le r\}$ for any $r>0$.

\end{lemma}

\no {\bf Proof.} We have  $|\ga_n(x)|=|x-2(s_1-s_2\dots)|\le x$,
then \er{211}, \er{212} give
$$
|y_n(x,\l)|\le \!\!\int_{G_n(x)}\!\! e^{|\Im \l|x}|V_{s_1}|\dots |V
_{s_n}|ds=\!\!\int_{G_n(x)}\!\! e^{|\Im \l|x}|v(s_1)|\dots
|v(s_n)|ds,
$$
since $|V (x)|=|v(x)|, x\in \T$. Then the standard estimates (see
\cite{PT87}) yield the first estimate in \er{217}, which shows that
for each $x>0$ the series \er{28} converges uniformly on bounded
subset of $\C\ts\mH$.  Each term of this series is an entire
function. Hence the sum is an entire function. Summing the majorants
we obtain estimates \er{217} and \er{218}. The proof of other results is standard (see e.g. \cite{PT87} or \cite{K08}) and  repeats the case for periodic $2\ts 2$ and
$2d\ts 2d$ Dirac operators, see \cite{K05}, \cite{K08} and we omit
it.
 $\BBox$

 \section {Traces of monodromy matrices and their properties \lb{Sec2x}}
\setcounter{equation}{0}

We describe the trace $T=\Tr \p(\l)$. At $v=0$ we have
$ \p_0(\l)=e^{-i\l J}$ and then $T_0=\Tr \p_0(\l)=e^{-i\l }+2e^{i\l }$.
 Using \er{211} we define the traces
\[
\lb{221}
\begin{aligned}
T_{2n}(\l):=\Tr\p_{2n}(\l)=\Tr \int_0^1\!\!\!\dots \!\!\!
\int_0^{s_{2n-1}}\!\!\!e^{i\l J\ga_{2n}}V_{s_1}\dots V_{s_{2n}}ds,
\end{aligned}
\]
where $\ga_{2n}=2(s_1-s_2+\dots -s_{2n})-1$ and
$s=(s_1,..,s_{2n})\in \R^{2n}$.  In particular, we have
\[
\lb{222}
\begin{aligned}
T_{2}(\l)=\!\! \Tr\int_0^1\!\!\int_0^{s_1}\!\!e^{i\l J(2(s_1-s_2)-1)} V_{s_1}V
_{s_2}ds=e^{-i\l}T_{2}^+(\l)+e^{i\l}T_{2}^-(\l),
\\
T_{2}^+=
\!\!\int_0^1\!\!\int_0^{s_1}\!\! e^{i2\l(s_1-s_2)} v^*(s_1)
v(s_2)ds,
\
T_{2}^-(\l)=\ol T_{2}^+(\ol \l)=\!\!\int_0^1\!\!\int_0^{s_1}\!\! e^{i2\l(s_2-s_1)} v^*(s_2)
v(s_1)ds.
\end{aligned}
\]
We sometimes write $y(x,\l,v),
\gD(\l,v),...$ instead of $y(x,\l), \gD(\l),,...$ when several
potentials  are being dealt with.

\begin{lemma}
\lb{Tmo} The function $T(\l,v)=\Tr \p(\l,v)$ is entire in $\l\in
\C$ and satisfies
\[
\lb{223} T(\l,v)=T(\l,-v)=\sum _{n\ge 0}T_{2n}(\l,v),
\]
\[
\lb{220} \Tr \p _{2n+1}(\l,v)=0,
\]
\[
\lb{223x}
\begin{aligned}
 T_{2n}(\l,v)=\int_{G_{2n}(1)}\!\! \Big(e^{i\l
\ga_{2n}}v^*(s_1)v(s_2)\dots v(s_{2n})+e^{-i\l
\ga_{2n}}\Tr v(s_1)v^*(s_2)\dots  v^*(s_{2n})  \Big)ds,
\end{aligned}
\]
\[
\lb{223xx} |T_{2n}(\l,v)|\le 2e^{|\Im \l|}{\|v\|^{2n}\/(2n)!},
\]
\[
\lb{224} |T(\l)|\le 4e^{|\Im \l|}\ch \|v\|,
\]
\[
\lb{226}
\begin{aligned}
 T(\l)=T_0(\l)+o(e^{|\Im \l|})\ \  \as \ \
|\l|\to \iy.
\end{aligned}
\]
Series \er{223} converge uniformly on bounded subsets of $\C\ts
\mH$. If a sequence $v^{(m)}$ converges weakly to $v$ in $\mH$ as
$m\to \iy$, then $T(\l,v^{(m)})\to T(\l,v)$ uniformly on bounded
subsets of $\C$.
\end{lemma}

 \no {\bf Proof.}
By Lemma \ref{Tfs}, series \er{223} converge uniformly and
absolutely on bounded subsets of $\C\ts \mH$. Each term in \er{223}
is an entire function of $z$, then $T$ is so.

\no Relations \er{213}-\er{215} yield   $\Tr V_{s_1}\dots V
_{s_{2n+1}}=0$ for any $s=(s_1,\dots s_{2n+1})\in \R^{2n+1}$. Then
$$
\Tr e^{i\l J\ga_{2n+1}(1,s)}JV_{s_1}\dots V_{s_{2n+1}}=0,
$$
for any $(x,s)\in \R\ts\R^{2n+1}$, which together with \er{212}
implies \er{220}.
From   \er{221} we have \er{223x}.

 We show \er{223xx}. Using
\er{223x}, $|\ga_{2n}|\le 1$ and $|V (x)|=|v(x)|$  we obtain
$$
|T_{2n}(\l,v)|\le 2\int_0^1\!\!\!\dots \!\!\! \int_0^{s_{2n-1}}\!\!
e^{|\ga_{2n}| |\Im \l|}|v(s_1)||v(s_2)\dots |v(s_{2n}|ds
$$
$$
\le 2e^{|\Im \l|}\int_0^1\!\!\!\dots \!\!\! \int_0^{s_{2n-1}}\!\!
|v(s_1)||v(s_2)|\dots  |v(s_{2n}|ds=2e^{|\Im \l|}
{\|v\|_1^{2n}\/(2n)!},
$$
which gives \er{223xx}. Using  \er{223} and \er{223xx} we obtain
\er{224}.  Asymptotics  \er{219} yields \er{226}.

If sequence $v^{(m)}$ converges weakly to $v$ in $\mH$ as $m\to
\iy$, then each $T_n(\l,v^{(m)})\to T_n(\l,v), n\in \N$ uniformly on bounded
subsets of $\C$. Using \er{211} and \er{212x} we obtain \er{223x}.
 $\BBox$

 We discuss the term $T_2$ more.

 \begin{lemma}
\lb{TT2} The function $T_2=\Tr \p_2=e^{-i\l}T_{2}^++e^{i\l}T_{2}^-$ has the properties
\[
\lb{t2b} T_{2}^++T_{2}^-=\wh v^*(\l)\wh {v}(-\l),\qqq T_2(\pi n)=(-1)^n|\wh v(\pi n)|^2,
\]
where $\wh v(\l)=\int_0^1\!\! e^{i2\l s} v(s)ds$ and $\wh {v^*}(\l)=\int_0^1\!\! e^{-i2\l s} v^*(s)ds$.
Moreover, let sequence $ |\l_n-\pi n|\leq
{\pi\/4}, n\in \Z$ and $\l_n\to \pm \iy$ as $n\to \pm \iy$. Then  for any fixed $\d>1$ the following asymptotics
\[
\lb{t2c}
\begin{aligned}
T(\l_n)=T_0(\l_n)+ T_2(\l_n)+\ell^{\d}(n) \qq   \as \ \
n\to \pm \iy
\end{aligned}
\]
hold, uniformly on $[0,1]\times\{|\l_n-\pi n|\leq
{\pi\/4}\}\ts \{\|v\|\le r\}$ for any $r>0$.

\end{lemma}

 \no {\bf Proof.} From \er{222} we obtain
$$
\begin{aligned}
T_{2}^-=\!\!\int_0^1\!\!\int_0^{s_1}\!\! e^{i2\l(s_2-s_1)} v^*(s_2)
v(s_1)ds
=\wh v^*(\l)\wh {v}(-\l)-
\!\!\int_0^1\!\!\int_{s_1}^1\!\! e^{i2\l(s_2-s_1)} v^*(s_2)v(s_1)ds
\\
=\wh {v^*}(\l)\wh {v}(-\l)-
\!\!\int_0^1\!\!\int_0^{s_2}\!\!e^{i2\l(s_2-s_1)} v(s_2)v(s_1)^*ds=\wh {v^*}(\l)\wh {v}(-\l)-T_{2}^+,
\end{aligned}
$$
which yields \er{t2b}. From \er{aspzn} we obtain \er{t2c}.
 \BBox

We determine asymptotics of the trace $T(\l)=\Tr \p(\l)$ as
$\l\in \mD_r, |\l|\to +\iy, r>0$, where the domain $\mD_r:=\{\l\in \C_+: \Im \l\ge r|\Re \l|\}.
$

\begin{lemma}
\lb{TasT} Let $\l=\x+i\n\in \mD_r$ and $|\l|\to \iy$ for some $r>0$.
Then
\[
\lb{aT1} T_2(\l)={1\/2\l } e^{-i\l }\big(\|v\|^2 +o(1)\big),
\]
\[
\lb{aT2} T_{4}(\l)={o(1)\/\l} e^{-i\l },
\]
\[
\lb{aT3} |T_n(\l)|\le {e^\n\/\n^{n\/4}} {\|v\|^{n}\/n!},\qq n\ge 4,
\]
\[
\lb{aT4} T(\l)=e^{-i\l } (1+{\|v\|^2+o(1)\/2\l } ).
\]

\end{lemma}

 \no {\bf Proof.}
From \er{222} and \er{TA-1}, \er{TA-1} we obtain
$$
\begin{aligned}
& T_2(\l)=\Tr\int_0^1dt\int_0^{x}e^{i\l J(2(x-s)-1)} V_xV_sds
\\
& =\int_0^1dx\int_0^{x} \rt(e^{-i\l } e^{i\l 2(x-s)} v^*(x)v(s)+\Tr
e^{i\l }e^{-i\l 2(x-s)}v(x)v^*(s)\rt)ds ={1\/2\l } e^{-i\l }(\|v\|^2
+o(1)).
\end{aligned}
$$
From \er{223x}, \er{TA-4} we have \er{aT2}.  From \er{223x} we have
for even $n\ge 6$:
\[
\lb{ee1} T_n(\l)=\!\!\int_{G_n(1)}\!\! \rt(e^{i\l
\ga_{n}}v^*(s_1)v(s_2)\dots v(s_{n})+e^{-i\l \ga_{n}}\Tr
v(s_1)v^*(s_2)\dots  v^*(s_{n})\rt)ds
\]
and due to \er{ab2} we have \er{aT3}. Substituting \er{aT1}-\er{aT3}
into \er{223} we obtain \er{aT4}.
 $\BBox$

 For a matrix-valued (and $\C$-valued) function $f(x,\l)$ we formally define the mapping by
$$
f(x,\l)\to \wt f(x,\l)=f^*(x,\ol \l),
$$
and   the Wronskian for the functions $f,g$:
\[
\{f,g\}=\wt fJg.
\]
If $f, g$ are solutions of the equation $iJ\p'+V \p=\l\p$, then the
Wronskian $\{f,g\}$ does not depends on $x$. In particular, the
Wronskian $\{\p,\p\}$ has the form
\[
\lb{3} \{\p,\p\}=\wt \p J\p=J \qq \Rightarrow \qq \p^{-1}=J\wt \p J.
\]
This identity is very usefull. For example, let  $\t=\t(\l)$ be a multiplier of $\p(\l)$. Note that
\er{4} implies $\t\ne 0$. Then the identity \er{3} yields
$$
\begin{aligned}
0=\det (\p(\l)-\t \1_3)=-\t^3\det \p(\l)\det (\p(\l)^{-1} -\t^{-1}\1_3)
\\
= -\t^3\det (J\wt\p(\l) J -\t^{-1} \1_3)= -\t^3\det (\p^*(\ol
\l)-\t^{-1} \1_3).
\end{aligned}
$$
Thus ${1/\t(\l)}$ is an eigenvalue of $\p^*(\ol \l)$ and ${1/\ol\t(\l)}$
is an eigenvalue of $\p(\ol \l)$. In particular, if $\l\in \R$,
then ${1/\ol\t(\l)}$ is an eigenvalue of $\p(\l)$.

 \no {\bf Proof of Theorem  \ref{T1}}. i) We show \er{asm} for
$\l\in \ol\C_+$, the proof for $\l\in \ol\C_-$ is similar. 
We have $\p=e^{-i\l J}+\p_{(1)}$, where $\p_{(1)}=\sum_{n\ge 1} \p_{n}$
and
\[
\lb{X1}
e^{-i\l J}=e^{-i\l}\os e^{i\l}\1_2,\qqq 
|\p_{(1)}(\l)|=e^{|\n|}o(1)\qq \as \qq |\l|\to \iy.
\]
 Then we have $\p=e^{-i\l}G$, where
\[
\lb{X2}
G=G_o+F,\qqq G_o=1\os e^{i2\l}\1_2,\qq F=e^{i\l}\p_{(1)}=o(1)\qq  \as \
|\l|\to \iy.
\]
$\bu $ Consider the case $\Im \l\ge r$ for $r>1$ large enough.
From \er{X2} and the perturbation theory for we obtain 
\[
\lb{X3}
\t_3(\l)=e^{-i\l}(1+o(1)),\qq \t_j(\l)=o(1)\qq \as \qq |\l|\to \iy.
\]
In order to determine asymptotics of $\t_j, j=1,2$ we
consider $\p(\l)^{-1}$ with the eigenvalues $\t_j^{-1}(\l), j\in
\N_3$.  From \er{3} and \er{219} we have
$$
\p(\l)^{-1}=J\p^*(\ol \l)J=e^{i\l J}+o(e^{|\Im \l|})\qq \as \qq
|\l|\to \iy.
$$
Using above arguments we obtain $\t_j(\l)=e^{-i\l}(1+o(1))$
as $|\l|\to \iy,\  \Im \l\ge r$ which yields 
\[
\lb{X4}
  \t_j(\l)=\t_j^o(\l)(1+o(1))\qq \as \qq |\l|\to \iy, \  \ \Im \l\ge r,
 \qq \forall \  j\in \N_3.
\]
$\bu $ Show asymptotics \er{asm} for the strip $S(r,\d)=\{\Im \l\in [0,r], |\l-\pi n|\ge \d, n\in \Z \} $. From \er{X2} we obtain
\[
\lb{X5}
G=\1_3+G_m+F,\qqq G_w=0\os 2ie^{i\l}\sin\l\1_2,\qq F=e^{i\l}\p_{(1)}=o(1)\qq  
\]
as $|\l|\to \iy, \ \l\in S(r,\d)$. For some $t>0$ small enough we have
$$
|F(\l)|\le t\qq \forall \ \l\in S(r,\d), |\Re \l|\ge m.
$$
for $m>1$ large enough.
From \er{aa1} we have 
\[
\lb{X6}
\d\sin \d\le |2ie^{i\l}\sin\l|\le 2 \qq \forall \l\in S(r,\d).
\]
We take $t<\d/N$ for $N$ large enough. Then by the perturbation theory
for matices we obtain
$$
 \t_j(\l)=\t_j^o(\l)(1+o(1))\qq \as \qq |\l|\to \iy, \ \l\in S(r,\d),
 \qq \forall \  j\in \N_3,
$$
which yields the needed results.

We show \er{5} for $D(\t,\l)=\det ( \p(\l)-\t\1_3)$. Let
 $w(x)=\det y(x,\l)$ for shortness. Due
to \er{Dd} for the determinant $w$ we have
$$
{w'(x)\/w(x)}=\Tr y'(x,\l)y(x,\l)^{-1}=\Tr iJ(V(x)-\l)=i\l ,
$$
since $\Tr JV_x=0$ and $\Tr J=-1$. Then gives $w(x)=e^{i\l x}$,
since $w(0)=1$.

 The direct calculation gives
$$
D(\t,\l)=-\t^3+\t^2\Tr \p(\l)+B(\l)\t +e^{i\l }, \qq\forall
\qq(\t,\l)\in\C^2,
$$
where $B(\l)={\pa\/\pa \t} D(0,\l)$. The identities \er{Dd}, \er{3}
and $D(0,\l)=e^{i\l }$ yield
$$
\begin{aligned}
B(\l)={\pa\/\pa \t} D(\t,\l)\big|_{\t=0}=-D(\t,\l)\Tr(
\p(\l)-\t\1_3)^{-1}\big|_{\t=0}= -e^{i\l }\Tr \p^{-1}(\l).
\\
=-e^{i\l }\Tr J\wt \p(\l)J=-e^{i\l }\Tr \wt\p(\l) =-e^{i\l }\ol
T(\ol \l),
\end{aligned}
$$
which yields \er{5}. We show \er{4}. We have $\wt T=\Tr
\p^{-1}={1\/\t_1}+{1\/\t_2}+{1\/\t_3}$. The direct calculation from
$\det (\p(\l)-\t \1_3)=-\t^3+\t^2T(\l)+e^{i\l }\t \wt
T(\l)+e^{i\l }$ gives
\[
\begin{aligned} \t_1+\t_2+\t_3=T,\qqq \t_1\t_2+\t_2\t_3+\t_1\t_3=e^{i\l }
\wt T(\l),\qq \t_1\t_2\t_3=e^{i\l }.
\end{aligned}
\]
ii)  We show that $\s(\cL)=\R$ by the contradiction. Assume that there is a gap $\g_o$ in the spectrum and then  $|\t_j(\l)|\ne 1$ for all $j\in \N_3, \l\in \g_o$.
Then due to ii) the numbers $1/\ol \t_j$ are also multipliers,
and we have at least 4 multipliers, which yields
a contradictions, since we have 3 multipliers.

Let $\l\in \R$. If all $|\t_j(\l)|=1, j=1,2,3$,
then this yields $\l\in \gS_3$.

If $|\t_1(\l)|>1$, then by ii) the number $\t_2={1/\ol\t(\l)}$ is a
multiplier of $\p(\l)$. Then the identity $\t_1\t_2\t_3=e^{i\l
}$ gives $|\t_3|=1$, which yields $\l\in \gS_1$.
\\
Let $|\t_1|=|\t_2|=1$. Then the identity $\t_1\t_2\t_3=e^{i\l }$
gives $|\t_3|=1$, which yields $\gS_2=\es$.
\BBox

We discuss the modified discriminant
$\gD=-{e^{-i2\l}\/64}(e^{ik_1}-e^{ik_2})^2
(e^{ik_1}-e^{ik_3})^2(e^{ik_2}-e^{ik_3})^2
$

\begin{lemma}  \lb{Tdd} Let multipliers $\t_j=e^{ik_j}, k_j\in\C, j\in \N_3$. Then
\[
\lb{rqm}
\begin{aligned}
\tes \gD=\sin^2{k_1-k_2\/2}\sin^2{k_1-k_3\/2}\sin^2{k_2-k_3\/2}.
\end{aligned}
\]
 If, in addition,  the quasimomentum $k_1=p_1+iq\in \C\sm \R, k_3\in \R$, then
\[
\lb{rqmg}
\begin{aligned}
& \tes {\gD}=-{1\/4}\sh^2{q}(\ch q-1+\sin^2 (p_1-k_3))^2<0.
\end{aligned}
\]
\end{lemma}

\no  {\bf Proof.} The definition \er{r} and the identity
$e^{i(k_1+k_2+k_3)}=e^{i\l }$ imply
$$
\begin{aligned}
& -64e^{2i\l
}\gD=(e^{ik_1}-e^{ik_2})^2(e^{ik_1}-e^{ik_3})^2(e^{ik_2}-e^{ik_3})^2
\\
& \tes =-64 \t_1^2\t_2^2\t_3^2\sin^2{k_1-k_2\/2}
\sin^2{k_1-k_3\/2}\sin^2{k_2-k_3\/2} =-64
e^{i2\l}\sin^2{k_1-k_2\/2}\sin^2{k_1-k_3\/2}\sin^2{k_2-k_3\/2}.
\end{aligned}
$$
 Let $k_1=p_1+iq$.  From $\t_2(\l)=1/ \ol\t_1(\l)$ we have that
$k_2=\ol k_1=p_1-iq$. This gives $k_1-k_3=a+iq$, and $k_2-k_3=a-iq$,
where $a=p_1-k_3$. Then from \er{rqm} we obtain
$$
\begin{aligned}
& \tes \gD=\sin^2{iq}\sin^2{a+iq\/2}\sin^2{a-iq\/2} =
-{\sh^2{q}\/4} (\ch q-\cos a)^2=-{\sh^2{q}\/4} (\ch q-1+\sin^2 a)^2<0,
\end{aligned}
$$
 which yields \er{rqmg}.
\BBox

\no  {\bf Proof of Theorem  \ref{Trho}.} i) Note  that the discriminant
$d$ of a polynomial  $-\t^3+\t^2a_2+a_1\t+a_0$ has the form
\[
\lb{disc} d=(a_1a_2)^2-4a_0a_2^3+4a_1^3-18a_0a_1a_2-27a_0^2.
\]
Applying this for the polynomial $D(\t,\l)$ given by
$
D(\t,\cdot)= -\t^3+\t^2T-\t e^{i\l } \wt
T+e^{i\l },
$
where $T=T(\l), \wt T=\ol T(\ol \l)$, we obtain
$$
\begin{aligned}
-64e^{2i\l }\gD(\l)=e^{2i\l }T^2\wt T^2-4e^{i\l }T^3- 4e^{3i\l }  \wt
T^3+18e^{2i\l }T \wt T-27e^{2i\l },
\end{aligned}
$$
which gives  \er{r}. Moreover,   the function $\gD(\l)$ is real for
real $\l$, since
\[
\lb{r+} -64\gD(\l)=|T(\l)|^4-4e^{-i\l }T^3(\l)-4e^{i\l } \ol
T^3(\l)+18|T(\l)|^2-27 \qqq \forall \ \l\in \R.
\]
We show \er{sro}. Let $\t_j=e^{ik_j}, k_j\in \C$  for $j\in \N_3$. Let all
$k_j(\l),j\in \N_3$ be different for all $\l\in\C$, except some finite
set on the unit interval. If the spectrum at $\l$ has multiplicity
3, then $ |e^{ik_j}|=1, j\in \N_3$, where each quasimomentum
$k_j(\l)\in\R$ and \er{rqm} gives $\gD(\l)\ge 0$.

If the spectrum at $\l$ has multiplicity 1, then only one from $k_1,
k_2, k_3$, (we denote it by  $k_3$)  is real. Then  from the identity \er{rqmg} we obtain $\gD(\l)<0$,  which yields
\er{sro}.

\no ii) If $v=0$, then it is clear that $\gD=0$.  Let $\gD=0$ for some $v\in\mH$. Then  due to asymptotics \er{asm} we get $\t_1=\t_2$. Thus the spectrum $\cL$ has the multiplicity 3. Then Theorem
\ref{Tk} yields $v=0$.
 \BBox

%\newpage
\section {Lyapunov functions \lb{Sec3}}
\setcounter{equation}{0}

In order to study the Lyapunov function we will do some
modification. Recall that if $\t$ is an eigenvalue of $\p(\l)$, then $1/\t$ is
an eigenvalue of $\wt \p(\l)$, since $\wt \p(\l)=J\p(\l)^{-1}J$.
We introduce the matrix-valued function
$\L={1\/2}(\p+\p^{-1})$ with eigenvalues
$\D_j={1\/2}(\t_j+\t_j^{-1}),j\in \N_3$. Due to \er{3} we obtain
\[
\lb{3.1}
\L(\l)={1\/2}(\p(\l)+\p^{-1}(\l))={1\/2}(\p(\l)+J\wt\p(\l)J).
\]
We discuss the properties of $\L(\l)$. Recall that  $T=\Tr \p(\l),
\wt T=\Tr \wt\p(\l)$.

%\newpage

\begin{lemma}
\lb{TL1} The determinant $D_\L(\l,\a):=\det (\L(\l,v)-\a \1_{3}),
\l,\a\in \C$ has the form
\[
\lb{L1}
\begin{aligned}
&   D_\L(\l,\a)=-\a^3+\a^2\cT(\l)-\a\cT_1(\l)+\det \L(\l),
\end{aligned}
\]
where
\[
\lb{L2} \tes  \cT=\Tr \L=\D_1+\D_2+\D_3={1\/2}(T+ \wt T),
\]
\[
\lb{L3} \tes   \cT_1=\D_1\D_2+\D_2\D_3+\D_1\D_3={1\/4}(e^{i\l }T+1)(e^{-i\l }\wt T+1)-1,
\]
\[
\lb{L4} \tes   \det \L=\D_1\D_2\D_3={1\/8}(2\cos \l +e^{i\l }\Tr
\wt\p^2+ e^{-i\l }\Tr \p^2),
\]
\[
\lb{L5} \Tr \p^2=T^2-2e^{i\l }\wt T,\qqq \Tr \wt
\p^2=(\wt T)^2-2e^{-i\l }T,
\]
where the coefficients $\cT, \cT_1, \det \L$ are real on the real
line. Moreover, the discriminant
$\vr_\D(\l)=(\D_1-\D_2)^2(\D_1-\D_3)^2(\D_2-\D_3)^2$ of the
polynomial $D_\L(\a,\l)$ in $\a\in \C$  is given by
\[
\lb{DL}
\begin{aligned}
\vr_\D(\l)=\cT^2\cT_1^2-4D_o\cT^3-4\cT_1^3-18D_o\cT\cT_1-27D_o^2,
\end{aligned}
\]
where $D_o=\det \L$.

\end{lemma}

\no {\bf Proof.} We have the standard decomposition $\det (\L(\l)-\a
\1_{3})=-\a^3+\cT \a^2- \a \cT_1 +\cD,$ where
$$
\begin{aligned}
 \cT=\D_1+\D_2+\D_3=\Tr \L,\qq \cT_1=\D_1\D_2+\D_2\D_3+\D_1\D_3,\qq
D_o=\D_1\D_2\D_3.
\end{aligned}
$$
From Theorem \ref{T1} we obtain
$$
\tes
\cT=\D_1+\D_2+\D_3={1\/2}(\t_1+{1\/\t_1}+\t_2+{1\/\t_2}+\t_3+{1\/\t_3})
={1\/2}(T+\wt T).
$$
In order to compute $\cT_1=\D_1\D_2+\D_2\D_3+\D_1\D_3$ we use similar arguments
$$
\begin{aligned}
\tes
\cT_1={1\/4}[(\t_1+{1\/\t_1})(\t_2+{1\/\t_2})+
(\t_2+{1\/\t_2})(\t_3+{1\/\t_3})+(\t_1+{1\/\t_1})(\t_3+{1\/\t_3})],
\end{aligned}
$$
and then
$$
\begin{aligned}
 4\cT_1=(\t_1\t_2+\t_2\t_3+\t_1\t_3)+
\rt({1\/\t_1\t_2}+{1\/\t_2\t_3}+{1\/\t_1\t_3}\rt)
+{\t_1\/\t_2}+{\t_2\/\t_1}+{\t_2\/\t_3}+{\t_3\/\t_2}+
{\t_3\/\t_1}+{\t_1\/\t_3}
\\
 =e^{i\l }\rt({1\/\t_1}+ {1\/\t_2}+{1\/\t_3} \rt)+e^{-i\l }({\t_1}+
{\t_2}+{\t_3})+({\t_1}+ {\t_2}+{\t_3})\rt({1\/\t_1}+
{1\/\t_2}+{1\/\t_3} \rt)-3,
\end{aligned}
$$
since ${\t_1\/\t_2}+{\t_3\/\t_2}+1={\t_1+
\t_2+\t_3\/\t_2}={T\/\t_2}$ and so on. This identity and Theorem
\ref{T1} give \er{L3}.

 Using $\t_1\t_2\t_3=e^{i\l }$
we compute the determinant
$$
\begin{aligned}
&  \tes \det
\L(\l)={1\/8}(\t_1+{1\/\t_1})(\t_2+{1\/\t_2})(\t_3+{1\/\t_3})
\\
& =\t_1\t_2\t_3+{1\/\t_1\t_2\t_3}+{\t_1\t_2\/\t_3}+
{\t_2\t_3\/\t_1}+
 {\t_1\t_3\/\t_2}+    {\t_1\/\t_2\t_3}+
{\t_2\/\t_1\t_3}+{\t_3\/\t_1\t_2}
\\
& \tes =e^{i\l }+e^{-i\l }+ e^{i\l
}\big({1\/\t_1^2}+{1\/\t_2^2}+{1\/\t_3^2} \big)+e^{-i\l
}\big(\t_1^2+\t_2^2+\t_3^2 \big)=2\cos \l +e^{i\l }\Tr \p^{-2}+
e^{-i\l }\Tr \p^2.
\end{aligned}
$$
The functions $\cT, \cT_1, \det \L$ are real on the real line, since
$
\wt T(\l)=\ol T(\l),
$
and \ $\cT_2(\l):=\Tr \wt \p^2(\l)=\Tr (\p^*(\l))^2=\ol
\cT_2(\l)$ for all $\l\in \R$. We have
$$
T^2=(\t_1+\t_2+\t_3)^2=(\t_1^2+\t_2^2+\t_3^2)+2(\t_1\t_2+\t_2\t_3+\t_3\t_1)
=\Tr \p^2+2e^{i\l }\wt T.
$$
Applying the mapping  $u\to \wt u$ to the last identity we obtain
$(\wt T)^2=\Tr \wt\p^2+2e^{-i\l }T$. Due to \er{L1} and \er{disc}
we obtain \er{DL}.
 \BBox

We discuss  properties of $\L$.
The identity \er{215} yields
\[
\lb{3.2}
\begin{aligned}
\L(\l)={1\/2}(\p(\l)+J\wt\p(\l)J)=\sum_{n\ge 0} \L_n(\l), \qqq
\L_n={1\/2}(\p_n+J\wt\p_nJ),
\end{aligned}
\]
where series in \er{3.2} converge uniformly on bounded subsets of
$\C\ts \mH$. Using \er{23},\er{211},\er{212} we obtain
\[
\begin{aligned}
\lb{3.3} J\p_{2n}^*(\ol \l)J=\int_0^1\! \!\!\dots \int
_0^{s_{2n-1}} e^{-i\l J\ga_{2n}}V_{s_{2n}}\dots
V_{s_1}ds=\p_{2n}^*(\ol \l)=\wt\p_{2n}(\l),
\\
 J\p_{2n+1}^*(\ol\l)J=-i\int_0^1\dots \int _0^{s_{2n}}
e^{-i\l J\ga_{2n+1}} V_{s_{2n+1}}\dots V_{s_1}Jds=-\p_{2n+1}^*(\ol
\l)=-\wt\p_{2n+1}(\l),
\end{aligned}
\]
where $\ga_n=2(s_1-s_2+...-(-1)^ns_{n})-1$, and in particular,
\[
\lb{3.4} J\p_1^*(\ol\l)J=-\p_1(\l),\qqq
J\p_2^*(\ol\l)J=\int_0^1\!\!\!\int _0^{s_1} e^{i\l
J(1-2s_1+2s_2)}V_{s_{2}}V_{s_1}ds_1ds_2.
\]
This yields     $\L_1=0$ and
$\L_n={1\/2}(\p_n(\l)+(-1)^n\p_{n}^*(\ol \l))$ and we deduce
that
\[
\begin{aligned}
\lb{36} \L_0(\l)=\1_3\cos \l  ,\qqq
\L_2(\l)={1\/2}\int_0^1\!\!\!\int _0^{s_1}\rt(e^{i\l J\ga_2}V_{s_1}V
_{s_2} +e^{-i\l J\ga_2}V_{s_{2}}V_{s_1}\rt)ds,
\\
\L_n(\l)={1\/2}\int_{G_n(1)}\rt(e^{i\l J\ga_n}V_{s_1}....V_{s_n}
+(-1)^ne^{-i\l J\ga_n}V (s_{n})...V_{s_1}\rt)ds,
\end{aligned}
\]
where $\L_{n}(\l)^*=(-1)^n\L_{n}(\l)$ for real $\l$ and $n\ge 2$.

\begin{lemma}  \lb{T31} Let $\mV=\int_0^1V_x^2dx$.
Then  the following asymptotics hold true as
$|\l|\to \iy$:
\[
\lb{asLm} \L(\l)=\cos \l  \1_3+ o(e^{|\Im \l|}),
\]
and if, in addition,
$\Im \l\ge r|\Re \l|$ for any fixed $r>0$, then
\[
\lb{37} \L_2(\l)=\rt({i\mV+o(1)\/2\l }\rt)\cos \l ,
\]
\[
\lb{381} \L(\l)=\rt(\1_{3}+{i\mV+o(1)\/2\l  }\rt)\cos \l ,
\]
\[
\lb{382} \det \L(\l)=(\cos^{3} \l) \exp\rt({i\|v\|^2+o(1)\/\l}\rt),
\]
\[
\begin{aligned}
\lb{asLF1} \D_j(\l)=\rt(1-{\gb_j+o(1)\/2i\l }\rt)\cos \l , \qq
j=1,2,3.
\end{aligned}
\]
\end{lemma}

 \no {\bf  Proof.} The asymptotics \er{asLm} follows from \er{219}.
Using $e^{iaJ}=\cos a+Ji\sin a$, we rewrite
$\L_2={1\/2}(\p_2(\l)+\wt \p_2(\l))$ in the form
\[
\lb{36z} \L_2(\l)={1\/2}\int_0^1\!\!\!\int _0^{s_1}\rt( K^+(s)\cos
\l\ga_2 +iJK^-(s)\sin \l\ga_2\rt)ds,\qq  s=(s_1,s_2)\in \R^2,
\]
where $K^\pm(s)=V(s_{1})V_{s_2}\pm V_{s_2}V_{s_1}$. Let $\l\in
\mD_r, |\l|\to \iy$. Due to \er{71} we obtain
\[
\int_0^1\!\!\!\int _0^{s_1}K^+(s)\cos \l \ga_2ds= {i\cos \l \/\l
}\rt(\int_0^1V^2_xdx+o(1)\rt),
\]
\[
\int_0^1\!\!\!\int _0^{s_1}K^-(s)\sin \l\ga_2ds= {i\cos \l \/2\l
}o(1),
\]
since $K^-(s)=0$ at $s_1=s_2$, and this yields \er{37}.

We estimate $\L_n, n>2$.
From \er{36} and \er{TA-4}   we have for $n=3,4,5$
\[
\lb{345} \L_n(\l)={1\/2}\int_{G_n(1)}\rt(e^{i\l J\ga_n}V_{s_1}....V
_{s_n} +e^{-i\l J\ga_n}V_{s_{n}}...V_{s_1}\rt)ds= {o(1)\/2\l }\cos
\l .
\]
and if we use \er{ab2} , then we obtain $ |\L_n(\l)|\le
{e^{\n}\/(4\x)^{{n\/4}}}{\|v\|^{n}\/n!} , \n=\Im \l$ and summing we
obtain
$$
|\sum_{n\ge 5} \L_n(\l)|\le {e^{\n}\/|\l|^{5\/4}}O(1),
$$
which jointly with \er{37}, \er{345}, \er{381} it   yields \er{381}.

Let $A={i \mV/2\l }$. Asymptotics \er{381}, \er{Das} give
\[
\det \L(\l)=\cos^3 \l\det(\1_{3}+A+o(1/\l))=\cos^3 \l(\1+\Tr
A+o(1/\l)),
\]
which yields  \er{382}. Asymptotics \er{asLF1} follows from
\er{381}, since the matrix $\L={1\/2}(\p+\p^{-1})$
has the eigenvalues $\D_j={1\/2}(\t_j+\t_j^{-1}),j\in \N_3$. \BBox

We discuss the identities for multipliers.

\begin{lemma}  \lb{TmL} i)
For distinct multipliers $\t_i, \t_j, \t_m$ the following identity holds true:
\[
\lb{mL1}
\begin{aligned}
&   (\t_i-\t_j^{-1})=(\t_m\t_j)^{-1}(e^{i\l }-\t_m),
\end{aligned}
\]
\[
\lb{mL2}
\begin{aligned} \D_i-\D_j={1\/2\t_i}(\t_i-\t_j)(\t_i-\t_j^{-1})={e^{-i\l}\/2}(\t_i-\t_j)(e^{i\l }-\t_m).
\end{aligned}
\]
\end{lemma}

 \no {\bf  Proof.} i) Using $\t_1\t_2\t_3=e^{i\l }$ we obtain  the identity
$$
(\t_i-\t_j^{-1})=\t_j^{-1}(\t_i\t_j-1)=(\t_m\t_j)^{-1}(\t_i\t_j\t_m-\t_3)
=(\t_j\t_m)^{-1}(e^{i\l }-\t_m).
$$
Properities of the multipliers give   the identity
$$
(\t_i-\t_j)(\t_i-\t_j^{-1})=\t_i(\t_i+\t_i^{-1}-\t_j-\t_j^{-1})=
2\t_i(\D_i-\D_j).
$$
Then jointly with \er{mL1} and the identity
$\t_1\t_2\t_3=e^{i\l }$ we obtain \er{mL2}.
\BBox

\no {\bf Proof of Theorem \ref{T2x}.} i) 
Asymptotics \er{asm}, \er{asLm}  give \er{LLa1}.
 From  asymptotics of the trace
formula \er{aT4} and the multipliers \er{asm} we have asymptotics of
$\t_1$ in \er{ras1}. Asymptotics \er{asm} and \er{asLF1} imply
asymptotics of $\t_2, \t_3$ in \er{ras1}. From \er{ras1} we have
$$
(\t_2(\l)-\t_1(\l))^2=e^{i2\l}(1-o(1)) \rt(e^{i{b_2+o(1)\/2\l
}}-e^{i{b_1+o(1)\/2\l }}
\rt)^2=-e^{i2\l}{(\gb_2-\gb_1)^2+o(1)\/4\l^2}.
$$
This jointly with \er{ras1} give
\[
\lb{}
\begin{aligned}
-64\gD(\l)=e^{-i2\l}(\t_1-\t_2)^2(\t_1-\t_3)^2(\t_2-\t_3)^2(\l)=
-e^{-i4\l}(1-o(1)){\gb_o+o(1)\/4\l^2},
\end{aligned}
\]
which yields \er{ras2}.

ii) If $\gb_1>0$, then  their definition gives $\gb_1<\gb_2<\gb_3$. Then
asymptotics of $\t_1,\t_2, \t_3$ in \er{ras1} yield that all multipliers are different and the Riemann surface is 3-sheeted.

iii) Recall that   the function
 $\gf=\cT_--\sin \l$, where $\cT_-={1\/2i}(T-\wt T)$.
From \er{5} we have
  $$
  e^{-i2\l }D(e^{i\l },\l)=-e^{i\l }+T-\wt T+e^{-i\l }=2i(\cT_--\sin \l)=2i\gf,
  $$
which yields \er{dd1}.  The identity \er{mL2} implies for $\t=e^{i\l }$:
$$
\begin{aligned}
&-(\t_1-\t_2)(\t_2-\t_3)(\t_3-\t_1)D(\t,\l)
\\
&  =(\t_1-\t_2)(\t_2-\t_3)(\t_3-\t_1)(\t-\t_1)(\t-\t_2)(\t-\t_3)=8e^{i3\l }(\D_1-\D_2)(\D_2-\D_3)(\D_3-\D_1),
\end{aligned}
$$
and then, using \er{dd1} we get
 $
-\gD(\l) D^2(e^{i\l },\l)=e^{i4\l }\vr_\D(\l)$, thus we get \er{dd2}.

Let $\cR$ be 3-sheeted and $\l_o\in \gS_1$ be a zero of $\gf$. Let some $|\t_m(\l_o)|=1, m\in \N_3$. Then due to \er{mL2} we obtain
$\t_i(\l_o)=\t_j(\l_o)$ or $\t_m(\l_o)=e^{i\l_o }$  for some
distinct multipliers $\t_i, \t_j, \t_m$.

If $\t_i(\l_o)=\t_j(\l_o)$, then all $|\t_j(\l_o)|=1$ and
 we have $\l_o\in  \gS_3$.

Let $\t_m(\l_o)=e^{i\l_o }$. Then the identity $\t_1\t_2\t_3=e^{i\l }$ gives
  $\t_i(\l_o)\t_j(\l_o)=1$ and $\D_i(\l_o)=\D_j(\l_o)$.
  We have two cases: $\D_i(\l_o)\in \R\sm [-1,1]$ or $\Im \D_i(\l_o)\ne 0$.

  1) $\Im \D_j(\l_o)\ne 0$: it is impossible since by Theorem \ref{T41},
  we have $\D_j(\l_o)=\ol \D_j(\l_o)$.

  2)  $\D_j(\l_o)\in \R\sm [-1,1]$: by Theorem \ref{T41}, the Riemann surface $\cR$  is 2-sheeted.

  Then all zeros of $\gf$ belong to $\gS_3$. From \er{dd2} and \er{sro}  we obtain  \er{dd3}, since
all zeros of $\gf$ belong to $\gS_3$.
\BBox

\begin{lemma}  \lb{Tpa}
Let $v\in \mH$.  Then there exists integer $n_*\ge 1$ such that
\\
i) the function $D_+$ has exactly $6(n_*+{1\/2})$ roots, counted
with multiplicity, in the disc $\{|\l|<2\pi(n_*+{1\/2})\}$ and
exactly 3 roots, counted with multiplicity, in each disk $\{|z-2\pi
n|<{1\/2}\}, |n|>n_*$. There are no other roots.
\\
ii) the function $D_-$ has exactly $6n_*$ roots, counted with
multiplicity, in the disc $\{|\l|<\pi n_*\}$ and  exactly 3 roots,
counted with multiplicity, in each disk $\{|z-\pi (2n+1)|<{1\/2}\},
|n|>n_*$. There are no other roots.

\end{lemma} \no {\bf Proof.} The proof repeats the proof for
periodic $2\ts 2$ and $2d\ts 2d$ Dirac operators, see \cite{K05},
\cite{K08} and we omit it.  $\BBox$

\no {\bf Proof of Theorem \ref{T3}.} Consider the case of even $n$,
the proof of odd case is similar.

\no i) We prove asymptotics \er{a8} for $z_{n,j},j\in \N_3$.
Firstly,  we prove rough asymptotics of $z_{n,j}$.
Let $D_+^o(\l)=2(1-\cos \l )(e^{-i\l }- 1)^2$ be
$D_+(\l)$  at $v=0$.
Using Lemmas
\ref{Tfs} and \ref{Tpa}, we obtain
$$
D_+(\l)=\det (e^{-i\l J}-\1_{3}+o(1))=D_+^o(\l)+o(1),\qq ,
$$ for
$|\l-2\pi n|<1, n\to \pm \iy $. Then the zeros of $D_+(\l)$ satisfy
\[
\lb{410} z_{n,j}=\pi n+x_{n,j}, \ \ x_{n,j}=o(1),\ \ j\in \N_3,\qq
n\to \pm \iy.
\]
Secondly, we improve these asymptotics. Due to \er{aspzn}, we have
the following asymptotics at $z_n=\pi n+x_n$ as $ n\to \pm \iy$(we
omit $j$ for shortness):
\[
\lb{ap}
\begin{aligned}
& iJe^{ix_nJ}(\p(\l_n)-\1_{3})=iJ(\1_3-e^{ix_nJ})-
\wh V(\l_n)+\ell^{\d/2}(n)
\\
& =x_n(1+O(x_n))-\wh V(z_n)+\ell^{\d/2}(n),
\\
& \wh V(\l)=\int_0^1 e^{i\l 2Js}V_sds,\qqq |u(z_n)|=
\ell^{\d}(n),\qq \forall \  \ \d>2,
\end{aligned}
\]
 which yields $x_n=\ell^\d(n)$.
We have $\wh V(\l_n)=\wh V(\pi n)+\wh V(\pi n)'x_n+O(x_n^2)$, where due to
\er{aspzn} $|\wh V(\pi n)'|=\ell^d(n)$. Repeating above arguments we
deduce that $|\wh V(\l_n)|=\ell^2(n)$ and $x_n=\ell^2(n)$. Hence due to
\er{ap}  we study the zeros of the equation
\[
\lb{ne} \det \big(\wh V(\pi n)+\ell^{\d/2}(n)-x_n\big)=0,\ \ \ x_n\in
\C.
\]
Let $\z_{n,j},j\in \N_3$ be the eigenvalues of the self-adjoint
operator $u(\pi n)$.   Using the arguments from the perturbation
theory for matrices (see [Ka,p.291]), we obtain that Eq. \er{ne} has
zeros $x_{n,j}=\z_{n,j}+\ell^{\d/2}(n),(j,n)\in \N_3\ts \Z$ which
yields \er{a8}. We compute $\z_{n,j}$. We have
$$
\begin{aligned}
\int_0^1 e^{i\pi n 2Jx}V_xdx=\ma   0 & \ol v_{1,n} & \ol v_{2,n}\\
            v_{1,n} & 0 & 0 \\
            v_{2,n} &  0 & 0 \am, \qq v_{s,n}=\int_0^1
           e^{i2\pi nx}v_s(x)dx,\ s=1,2,
\end{aligned}
$$
This gives that $\z_{n,3}=0, \z_{n,s}=\pm |\wh v(\pi n)|$, where
$\wh v(\l)=\int_0^1 e^{i2\l x}v(x)dx$.

We show \er{Dpm1} for $D_+$, the proof for $D_-$ is similar.
Identity \er{5} gives at $\t=1$:
\[
\lb{D+D-}
\begin{aligned}
\tes D_+=(T-1)-e^{i\l} (\wt T-1)=ie^{i{\l\/2}}F_+, \qq
F_+={1\/i}[e^{-i{\l\/2}}(T-1)-e^{i{\l\/2}} (\wt T-1)],
\\
D_-=(T+1)+e^{i\l} (\wt T+1)=e^{i{\l\/2}}F_-, \qq
F_-=[e^{-i{\l\/2}}(T+1)+e^{i{\l\/2}} (\wt T+1)].
\end{aligned}
\]
The function $F_+$ is entire and real on the real line. Due to
\er{asm} $F_+$ has the types $\gt_\pm(F_+)={3\/2}$ and due to
\er{esT} the function $F_+$ is bounded on the real line. This gives
$F_+\in E_{Cart}$ and \er{Dpm1}, since $F_+(0)={1\/i}(T(0)-\ol
T(0))=2\Im T(0)$.

 Taking a sum and the difference in \er{D+D-} we obtain \er{fjd1}.
Using the identity $-64\gD=T^2\wt T^2-4e^{-i\l }T^3-4e^{i\l } \wt
T^3+18T\wt T-27$ we recover $\gD$ and the Riemann surface. \BBox

\section {Conformal mappings and trace formulas \lb{Sec4}}
\setcounter{equation}{0}

\subsection{Quasimomentums}
Define a conformal mapping $\e:\C\sm [-1,1]\to \{\z\in \C: |\z|>1\}$ by
\[
\lb{dx} \e(z)=z+\sqrt{z^2-1},\qq z\in \C\sm [-1,1],\ \ {\rm
and}\qq \e(z)=2z+o(1),\ \ |z|\to \iy.
\]
Note that $\e(z)=\ol \e(\ol z), z\in \C\sm [-1,1]$ since $\e(z)>1$
for any $z>1$.  Let $\B_f$ be the set of all branch points of the
function $f$.  Define the domains
$$
\cR_\pm=\{\z\in\cR:\pm\Im\z>0\},\qq \cR_\pm^o=\C_\pm\sm\b_\pm,\qqq
\cR^o=\C\sm (\b_+\cup \b_-\cup \b_0),
$$
$$
\b_\pm={\bigcup}_{\b\in\B_\D\cap\C_\pm}[\b,\b\pm i\iy), \qqq
\b_0=\{\l\in \R: \D_j(\l)\notin \R \ for \ some \ j\in \N_3\},
$$
where $\cR_\pm^o$ are simply connected.
 Due to \er{dsp}  we have
$\D(\z)\notin [-1,1]$ for $\z\in\cR_+$. For each
$\D_j={1\/2}(\t_j+\t_j^{-1})$ we introduce the quasimomentum (some
branch)
\[
\lb{dkj} k_j(\z)=\arccos \D_j(\z)=i\log \e(\D_j(\z)),\ \ \ \ \z\in
\cR_+,\ \ \  j\in \N_3.
\]
The Lyapunov function $\D(\z)$ is analytic on some $3$--sheeted
Riemann surface $\cR$ and $q(\z)=|\log\e(\D(\z))|$ is the
single-valued function on $\cR_+$ and is the imaginary part of the
(in general, many-valued on $\cR_+$) quasimomentum $k(\z)$ given by
$$
k(\z)=p(\z)+iq(\z)=\arccos\D(\z)=i\log\e(\D(\z)).
$$
We denote by $q_j(\l)$, $(\l,j)\in\C_+\ts \N_3$, the branches of
$q(\z)$ and by $p_j(\l)$, $k_j(\l)$, $\l\in\cR_+^o$, the
single-valued branches of $p(\z)$, $k(\z)$, respectively.
Let $\dP: \cR\to \C$ be the natural projection of $\z\in\cR$ onto $\C$
given by $\dP(\z)=\l=\x+i\n\in\C$. We describe Lyapunov functions $\D$ on
Riemann surface.

\begin{theorem}
\lb{T41}
The function $q(\z)=\log |\e(\D(\z))|$ is
subharmonic on the Riemann surface $\cR$ and has the following
asymptotics:
\[
\lb{41r}
 q(\z)=\n+O(|\l|^{-{1\/2}})\qq \as \qq |\l|\to \iy, \  \z\in
\ol\cR_+.
\]
Let, in addition,  $\g_o=(\l^-, \l^-)$ be some open gap in the
spectrum $\gS_3$ and let an interval $\o\ss \g_o$, and $\D_3(\o)\ss
[-1,1]$ for some labeling.

\no i) If some $\D_1(\l)\in \R\sm [-1,1]$ for all $\l\in \o$, then
the Riemann surface $\cR$ is 2-sheeted and
\[
\lb{idz}
\begin{aligned}
&  \t_1\t_2=1, \qq \D_1=\D_2, \qq \t_3=e^{i\l},
\\
& D(\t,\cdot)=-(\t^2-2\t\D_1+1)(\t-e^{i\l}),\qq
%\vr=16(\D_{1}^2-1)(\D_{1}-\cos \l )^2,
\gD={1\/4}(1-\D_{1}^2)(\D_{1}-\cos \l )^2
\end{aligned}
\]
and $k_1(\cdot)$ has an analytic extension from $\cR_+^o$ into
$\cR_\o=\cR_+^o\cup\cR_-^o\cup \o$ such that
\[
\lb{43}
\begin{aligned}
\Re k_1(\l)={\rm const}\in \pi \Z,\ \ \forall \ \  \l\in \o,
\\
q_1(\l)=q_1(\ol \l)>0,\ \ \ \ \l\in \cR_\o.
\end{aligned}
\]
 ii) Let $\D_1(\l)$ be complex  on $\o$. Then  the Riemann surface $\cR$
is 3-sheeted; the functions $\D_1(\l)$ and $k_{1}+k_2$ have analytic
extensions from $\cR_+^o$ into $\cR_{\o}=\cR_+^o\cup \cR_-^o\cup \o$
such that
\[
\lb{45a4} \D_1(\l)=
\ca \D_1(\l)\ \ \ &if\ \ \ \ \l\in \cR_+^o,\\
       \ol\D_{2}(\ol \l)\ \ \ &if\ \ \ \ \l\in \cR_-^o,\ac
\]
\[
\lb{464}
\begin{aligned}
\ca \ p_1(\l)+p_{2}(\l)={\rm const}\in 2\pi \Z,
 \\
\  0<q_1(\l)=q_{2}(\l)\le \|v\|,\ \ \ \ \ac
\end{aligned}
\qqq \l\in \g_o,
\]
\[
\lb{474}
 q_1(\l)+q_{2}(\l)=q_1(\ol \l)+q_{2}(\ol \l)>0,\ \
\ \l\in \cR_{\g_o}.
\]
\end{theorem}

\no {\bf Proof.} Results about $q$ and \er{41r} were proved in
\cite{K10} and we need to show i) and ii) only.
\\
i) If $\D_1(\o)\ss \R\sm [-1,1]$, then
$\D_1(\l)={1\/2}(\t_1(\l)+\t_1(\l)^{-1})$, where $\t_1(\l)$ is real
on $\o$. Thus due to Theorem \ref{T1}, i) the multiplier
$\t_2(\l)=1/\t_1(\l)$ is real, and then $\D_1(\l)=\D_2(\l)$ for all
$\l\in \g_o$, which yields $\D_1=\D_2$ and $\t_2\t_1=1$, since these
functions are analytic on $\cR$.

From  $\t_1\t_2\t_3=e^{i\l}$, we immediately have $\t_3=e^{i\l}$,
and from \er{5}, \er{r0} we obtain \er{idz}.
\\
 ii) It was shown in \cite{K10} that the functions $\D_1(\l)$ and
$k_{1}+k_2$ have analytic extensions from $\cR_+^o$ into
$\cR_{\o}=\cR_+^o\cup \cR_-^o\cup \o$ and satisfy \er{45a4}-\er{464}
(with the exception of the estimate  $q_1(\l)\le \|v\|$ in
\er{464}). Thus we have two complex different functions $\D_1,
\ol\D_1$ and one real function $\D_3$ on $\o$. Thus the Riemann
surface $\cR$ is 3-sheeted. We need to estimate  $q_1(\l)$ on $\o$.

Let $\l\in \o\ss \gS_1$. Then $|\t_3(\l)|=1$ and  from  \er{4} we
have  $|\t_1(\l)||\t_2(\l)|=1$. We assume that
$|\t_1(\l)|<1<|\t_2(\l)|$, the proof for another case is similar.
Then from \er{217} we have $|\t_2(\l)|\le |\p(\l)|\le e^{\|v\|}$,
which yields $e^{q_2(\l)}\le e^{\|v\|}$ and  $q_2(\l)\le \|v\|$.
\BBox

 In Theorem \ref{T41}, ii) we have
$\D_2=\ol\D_1$ on $\g_o$. Thus we have 3 different functions $\D_j,
j\in \N_3$ and then the Riemann surface $\cR$ is 3-sheeted. It is very difficult to determine
the positions of branch points.

\subsection{Conformal mappings}

A real function $f$ on the real line is upper semi-continuous if,
for any $p,r\in \R$ satisfying $f(p) < r$, there is a neighborhood
$U\ss\R$ of $p$ such that $f(p_1) < r$ for every $p_1 \in U$. Let
$C_{us}$ denote the class of all real upper semi-continuous
functions  $h:\R\to \R$. With any $h\in C_{us}$  we associate the
"upper" domain
$$
\K(h)=\Big\{k=p+iq\in\C: q>h(p), p\in \R\Big\}.
$$
Define an {\it averaged quasimomentum}  $\dk$ (shortly AK), a real function $\gp$ and a {\it Lyapunov exponent} $\gq$ by
\[
\lb{dak} \dk(\l)=\gp(\l)+i\gq(\l)={1\/3}\sum_1^3 k_j(\l), \ \ \  \gq(\l)=\Im
\dk(\l), \ \  \l\in \cR_+^o.
\]
We present results about conformal mappings and identities from \cite{K10}, which will be important to estimate spectral parameters for VNLSE, similar to the KdV equation, see \cite{KK95}
\cite{K98}, \cite{K97}, \cite{K06}   and for the scalar NLSE \cite{K96}, \cite{K01}, \cite{K05}.

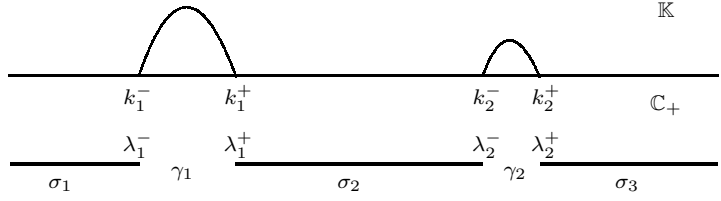
\begin{figure}
\tiny
\unitlength 0.7mm % = 2.845pt
\linethickness{0.4pt}
\ifx\plotpoint\undefined\newsavebox{\plotpoint}\fi % GNUPLOT compatibility
\begin{picture}(138.675,40.737)(0,0)
\put(5.325,26.75){\line(1,0){133.35}}
\qbezier(29.825,26.75)(38.575,52.737)(48.025,26.575)
\qbezier(94.4,26.75)(99.212,40.138)(105.075,26.575) \thicklines
\put(5.675,10.){\line(1,0){24.15}}
\put(48.025,10.){\line(1,0){46.375}}
\put(105.25,10.){\line(1,0){33.25}}
\put(128.875,39.175){\makebox(0,0)[cc]{$\K$}}
\put(128.875,21.175){\makebox(0,0)[cc]{$\C_+$}}
\put(29.65,22.55){\makebox(0,0)[cc]{$k_1^-$}}
\put(48.55,22.55){\makebox(0,0)[cc]{$k_1^+$}}
\put(95.275,22.55){\makebox(0,0)[cc]{$k_2^-$}}
\put(106.3,22.55){\makebox(0,0)[cc]{$k_2^+$}}
\put(29.65,13.625){\makebox(0,0)[cc]{$\l_1^-$}}
\put(48.55,13.625){\makebox(0,0)[cc]{$\l_1^+$}}
\put(95.275,13.625){\makebox(0,0)[cc]{$\l_2^-$}}
\put(106.3,13.625){\makebox(0,0)[cc]{$\l_2^+$}}
\put(15.125,6.1){\makebox(0,0)[cc]{$\s_1$}}
\put(69.375,6.1){\makebox(0,0)[cc]{$\s_2$}}
\put(121.525,6.1){\makebox(0,0)[cc]{$\s_3$}}
\put(38.05,8.2){\makebox(0,0)[cc]{$\g_1$}}
\put(100.525,8.2){\makebox(0,0)[cc]{$\g_2$}}
\end{picture}
\lb{fig3} \caption{\footnotesize The domain $\K$ and
$k_n^\pm=k(\l_n^\pm)$ }
\end{figure}

\begin{theorem}
\lb{Tk} Let $\dk={1\/3}\sum_1^{3} k_j$ be the averaged quasimomentum.
Then
\\
i) The function $\dk(\cdot)$ is analytic in $\C_+$, and  is a
conformal mapping from $\C_+$ onto $\dk(\C_+)=\K(h)$  for some $h\in
C_{us}$. Moreover, $\gq=\Im \dk>0$ for any $\l\in \C_{+}$.
\\
ii) There exist branches $k_j,j\in \N_3$, such that the following
relations hold true:
 \[
 \lb{T23-1}
\dk(\l)=\l+{1\/\pi}\int_{\R}{\gq(t)\/t-\l}dt, \ \ \ \l=\x+i\n\in\C_+,
\]
 \[
 \lb{T23-2}
\dk(\l)=\l-{Q_o+o(1)\/\l}
\]
as $ \l\in \mD_r$, $ |\l|\to \iy$,  for any $ r>0$, and
 \[
 \lb{T23-3}
Q_o={1\/\pi}\int_{\R}\gq(\x)d\x={1\/\pi}\iint_{\C_+}|\dk'(\l)-1|^2d\n
d\x+{1\/\pi}\int_{\R}\gq(\x)d\gp(\x)={2\/3}\|v\|^2,
\]
\[
 \lb{T23-4}
\tes
 \gq|_{\gS_3}=0,\qqq 0<\gq|_{\gS_1}\le\|v\|= ({2\/3}Q_o)^{1\/2}.
\]
iii) Let $v\in \mH$.  Then we have
  $\gS_3=\R$ iff $v=0$.

\end{theorem}

\no {\bf Proof.}
The theorem is proved for more general case from \cite{K10}.
We need only to show  $\gq|_{\gS_1}\le ({2\/3}Q_o)^{1\/2}$  in \er{T23-4},
since all other results were proved in \cite{K10}.
From  \er{464} we have $\gq(\l)={q_1(\l)+q_2(\l)\/3}={2q_1(\l)\/3}$
for any $\l\in \g_n$. This and \er{T23-3}  imply $\gq|_{\gS_1}\le
({2\/3}Q_o)^{1\/2}$.
 \BBox

\no {\bf Remark.} 1) The integral ${1\/\pi}\int_{\R}\gq(\x)d\gp(\x)\ge 0$ is the area between the boundary of $\K(h)$  and the real line.

\no 2) It was shown in \cite{K10} that for any AK
$\dk=\gp+i\gq$ the function $\gp'_\m(\l)>0, \l=\x+i\n\in {\C}_+$. Hence
there are two positive functions $\gq$ and $\gp'_\x$ in $\C_+$. From
\er{T23-1} we have
\[
\gq(\l)=\n+{\n\over \pi}\int {\gq(t)\/ |t-\l|^2}dt, \ \ \ \l\in
{\C}_+,
\]
 \[
 \lb{T3-1}
\gp'(\l)=1+{1\/\pi}\int_{\R}{\gq(t)\/(t-\l)^2}dt\ge 1, \ \qq \l\in
\gS_3,
\]
here $\gp_\x'(\l)=1$ for some $\l\in\gS_3$ iff $\gS_3=\R$.

\subsection{Riemann surfaces }

We discuss Lyapunov functions on 2-sheeted Riemann surfaces.

\no {\bf Proof of Theorem \ref{T2sR}.}
 We show \er{uzs2}, the proof of \er{uzs3} is similar.
 We assume that $v\ne 0$. Let $\g_o\ss \gS_1$ be
some open gap in the spectrum $\gS_3$ for some $v\ne 0$ and let an interval $\o\ss
\g_o$. We show that following conditions (1),..,(5)  are equivalent
\[
\lb{c14}
 (1) \ \ \D_2(\o)\ss \R\sm [-1,1] \ (2)\ \ \D_2=\D_3, \ (3)\ \
\t_2\t_3=1, \ \ (4)\ \ \t_1=e^{i\l}, \ \ (5)\ \
\ \ \cR\in \mR_2.
\]
We show $(1)\Rightarrow (2)$. Let $\D_2(\o)\ss \R\sm [-1,1]$. Then $\D_2(\l)={1\/2}(\t_2(\l)+\t_2(\l)^{-1})$, where $\t_2(\l)$ is
real on $\o$. Thus due to Theorem \ref{T1}, i) the multiplier
$\t_3(\l)=1/\t_2(\l)$ is real. Note that the case $\t_1(\l)=1/\t_2(\l)$ is impossible, since we have asymptotics \er{asm}. Thus $\D_2(\l)=\D_3(\l)$ for all
$\l\in \o$, which yields $\D_2=\D_3$, since these functions are
analytic on $\cR$.

 $(2)\Rightarrow (3)$. Let $\D_2=\D_3$ on $\o$.
Due to Theorem \ref{T41} the function $\D_2$ on $\o$ can be real or
complex. Then we have $\D_2=\D_3$ are real  on $\o$. At this
condition we have proved $\t_2(\l)\t_3(\l)=1$ on  $\o$.

$(3)\Rightarrow (4)$. Let $\t_1(\l)\t_2(\l)=1$ on  $\o$. From
$\t_1\t_2\t_3=e^{i\l}$, we immediately have $\t_3=e^{i\l}$, which
yields $ (4)$.

$(4)\Rightarrow (1)$. Let $\t_1=e^{i\l}$. From
$\t_1\t_2\t_3=e^{i\l}$, we have $\t_2\t_3=1$ on $\o$ and we again
have that $\D_2=\D_3$ are real  on $\o$, since $\o\ss \gS_1$.

$(5)\Rightarrow (4)$.
Let $\cR\in \mR_2$, i.e., it is the 2-sheeted Reiemann surface. Due to Theorem
\ref{Tk}, iii) there exists an open gap $\g_o\ss \gS_1$ and
$\D_j(\g_o)\ss [-1,1]$ for some $j\in \N_3$. We have three cases:

$\bu$ Let $\D_1(\g_o)\ss [-1,1]$.
Then due to Theorem \ref{T41} the function
$\D_2(\l)$ is real or complex on $\o\ss \g_o$. If $\D_2(\l)$ is
complex on $\g_o$, then due to Theorem \ref{T41} we have 3 functions
$\D_2(\l)=\ol \D_3(\l)$ and $\D_1(\l)\in [-1,1]$ on $\g_o$ and  we
have 3 different analytic functions on $\cR$,  and $\cR$ is
3-sheeted. Then the function $\D_2$ is real on $\o$ and   $\D_2(\o)\ss \R\sm [-1,1]$, which is
 equivalent to one from the  conditions (1),..,(4).

$\bu$ The condition  $\D_1(\o)\ss [-1,1]$ corresponds to the second case
\er{uzs3}.

$\bu$ Let $\D_3(\o)\ss [-1,1]$.
Then due to Theorem \ref{T41} the function
$\D_2(\l)$ is real or complex on $\o\ss \g_o$. If $\D_2(\l)$ is
complex on $\g_o$, then due to Theorem \ref{T41} we have 3 functions
$\D_2(\l)=\ol \D_3(\l)$ and $\D_1(\l)\in [-1,1]$ on $\g_o$. Thus we
have 3 different analytic functions on $\cR$,  and $\cR$ is
3-sheeted. Then the function $\D_2(\o)\ss \R\sm [-1,1]$, which is
 equivalent to one from the  conditions (1),..,(4).

 $(5)\Rightarrow (4)$.
Conversly, let $\t_1=e^{i\l}$. Then  $\t_2\t_3=1$, which yields
$D(\t,\l)=-(\t^2-2\t\D_2+1)(\t-e^{i\l})$. Then $\cR\in \mR_2$ and
conditions (1),..,(5) in \er{c14} are equivalent.

In order to finish the proof we show  that following conditions (5),..,(9) in \er{qq1} are equivalent
\[
\lb{qq1}
\begin{aligned}
(5)\ \cR\in \mR_2  \Leftrightarrow  (6)\   v\in \mH_o \   \Leftrightarrow
\ (7)\   \gf=0 \  \Leftrightarrow \ (8)\ \vr_\D=0  \ \Leftrightarrow \ (9)\   \D_3(\o)\ss \R \sm[-1,1].
\end{aligned}
\]
Show $(5) \Leftrightarrow  (6)$.
 Let $\cR\in \mR_2$ for some $v\in \mH$.
Then by \er{c14}, the multiplier $\t_1=e^{i\l}$ and $\t_2\t_3=1$. Asymptotics
\er{ras1} gives
$$
\gb_1=0 \Leftrightarrow c_{12}=|\int_0^1 v_1\ol v_2dx|=\|v_1\|\|v_2\|.
$$
Thus from this identity we obtain $v=ue\in \mH$  for some $(u,e)\in \mH\ts \C^2$. Conversely, let $v\in \mH_o$. Then
Proposition \ref{Tpr} gives that the corresponding Riemann surface $\cR\in \mR_2$.

Show $(6) \Leftrightarrow  (7)$.
If $v\in \mH_o$, then due to \er{mtt} we have $\t_1=e^{i\l}$ and the identity
$D(e^{i\l},\l)=2ie^{i2\l}\gf(\l)$ from \er{dd1} gives $\gf=0$.
Conversely, let $\gf=0$. Then the identity
$D(e^{i\l},\l)=2ie^{i2\l}\gf(\l)$ give that $\t_1=e^{i\l}$ (or the second case $\t_2=e^{i\l}$ )
since $\t_3=e^{-i\l}$ as $\Im \l\to +\iy$.

Show $(7) \Leftrightarrow  (8)$.
From the identity $ 4\gD \ \gf^2=\vr_\D$ (see \er{dd2}) we deduce that
$\vr_\D=0\Leftrightarrow \gf=0$, since by Theorem \ref{Trho}, $\gD\ne 0$ for $v\ne 0$.

$(1)  \Leftrightarrow   (9)$. Let $\D_3(\o)\ss \R\sm [-1,1]$. Then $\D_3(\l)={1\/2}(\t_3(\l)+\t_3(\l)^{-1})$, where $\t_3(\l)$ is
real on $\o$. Thus due to Theorem \ref{T1}, i) the multiplier
$\t_2(\l)=1/\t_3(\l)$ is real (or  $\t_1(\l)=1/\t_3(\l)$ the second case), and then $\D_2(\l)=\D_3(\l)$ for all
$\l\in \o$, which yields $\D_2=\D_3$ and  (1) in \er{c14}, since these functions are analytic on $\cR$. The proof of $(9) \Rightarrow  (1)$ is similar.
\BBox

\no {\bf Proof of Corollary \ref{T6}.} From Theorem \ref{Trho} and \ref{Tk}
we have
\[
\lb{61q}
\begin{aligned}
v=0\qq \Leftrightarrow \qq  \gS_3=\R \qq  \Leftrightarrow
\qq \gD=0.
\end{aligned}
\]
We need to consider $\D_1=\D_2$ and $ \t_1=\t_2$. It is clear that
if $v=0$, then $\D_1=\D_2$ and $ \t_1=\t_2$.

Let $\D_1=\D_2$. Then $\vr_\D=0$ and the identities \er{dd2}, \er{dd3} imply that $\gD\gf=0$. If $\gD=0$, then \er{61q} gives $v=0$. Let $\gf=0$.
Then due to \er{dd3} we have  $D(e^{i\l},\l)=0$ and then $e^{i\l}=\t_1$
and by $\D_1=\D_2$,  we obtain $e^{i\l}=\t_1=\t_2$. And finally
the identity $\t_1\t_2\t_3=e^{i\l}$ gives $\gD=0$.

Let $\t_1=\t_2$. Then we have $\D_1=\D_2$, which have been discussed above.
\BBox

We describe properties of the Lyapunov functions including large
$n\in \Z$. For small $\ve
>0$ and  $n\in \Z$ large enough due to \er{a8} we define sets
$$
I_n:=[\pi (n-1)+\ve, \pi n-\ve], \qq \cZ_n:=\{z_{n,1}\le z_{n,2}\le
z_{n,3}\}\ss g_n:=[\pi n-\ve, \pi n+\ve].
$$

\begin{theorem}
\lb{Tbp} Let  the Riemann surface $\cR$ be 3-sheeted.
\\
i) Let $\D$ have a branch point $\l_o\in \R$ such that that each
$\D_j(\l)\in \R, j\in \N_3$, for all $\l\in (\l_o,\l_o+\ve)$ (or all
for $\l\in (\l_o-\ve,\l_o)$) for small $\ve>0$. Then $\l_o$ is a
branch point of order ${1\/2}$ for $\D_1, \D_2$, and they have
asymptotics as $ t:=\l-\l_o\to 0$:
\[
\lb{bpas} \D_1(\l)=\D_1(\l_o)+C\sqrt{t}(1+O(t)),\ \
\D_2(\l)=\D_1(\l_o)-C\sqrt{t}(1+O(t))\qq
\]
for some constant $C$, where $\sqrt 1=1$ (and similar asymptotics
for real $\D_j$ for $\l\in (\l_o-\ve,\l_o)$). The function
$(\D_1-\D_2)^2$ is analytic in the disk $\{|\l-\l_o|<\ve\}$ and it
has a zero $\l_o$ of multiplicity $2m+1\ge 1$. If in addition
$\D_1(\l_o)\in (-1,1)$, then $m=0$.
\\
ii) Let $n\in \Z$ be  large enough. Let $\D_1$ be real on some
interval $\o=(\a,\b)\ss I_n$. Then $\D_1$ is real analytic on the
interval $I_n$.

\end{theorem}
\no {\bf Proof.} The statement i) was proved in  \cite{K10}.
\\
ii) Let   $n\in 2\Z$ and let $\l_o\in I_n$ be a branch point of
$\D_1$ such that $\l_o\le \a$. The proof of other cases is similar.
Then asymptotics \er{LLa1} gives that all $\D_j(\l_o)\in (-1,1)$ and
the statement i) gives that $\l_o$ is a branch point of the order
${1\/2}$ for $\D_1, \D_2$ and has asymptotics \er{bpas} as $
t:=\l-\l_o\to 0$ (for some labeling). Then the Lyapunov function
$\D$ has a loop and its branch has a local maximum $|\D_2(z_o)|<1$
at some point $z_o>\l_o$. This contradict to the monotonicity
property of the Lyapunov function and thus $\D$ does not have branch
points on $I_n$. \BBox

We discuss the MS for two sheeted Riemann surfaces.
%\newpage

\begin{lemma} \lb{Tcom} i) Let $v$ be a solution of the  MS. Then each intergal
$c_{ij}=\int_0^1 v_i(x,t)\ol v_j(x,t)dx, i,j=1,2$ does not depend on time,
i.e., is a constant of motion. Moreover, if $v_1=0$ or $v_2=0$, then the  MS  $i\dot v=-v''+|v|^2v$ is the scalar NLS equation.
\\
ii) Let $v_1  \bot v_2$. Then the coefficients $\gb_1, \gb_2, \gb_3$ from \er{V23xx} are given by
\[
\lb{gb1}
\begin{aligned}
%\tes
\gb_3=\|v\|^2,\qq \gb_2=\max \{\|v_2\|^2, \|v_1\|^2\}, \qq
\gb_1=\min \{\|v_2\|^2, \|v_1\|^2\},
\end{aligned}
\]
\[
\lb{gb2}
{\rm if} \qq     \cR\in \mR_2\qq \Rightarrow \qq
v_1=0 \ \ {\rm or}\ \ v_2=0.
\]

\end{lemma}
\no {\bf Proof.} i) Let $\lan \cdot, \cdot\ran $ be the inner product in $L^2(\T)$.
Let $\dot y={\pa \/\pa t}y$ and $y'={\pa \/\pa x}y$. Then
$$
\begin{aligned}
i\dot c_{12}=i\lan \dot v_1, v_2\ran+i\lan \dot v_1, \dot v_2\ran
=\lan -v_1''+|v|^2v_1, v_2\ran+\lan v_1, v_2''-|v|^2v_2\ran=0.
\end{aligned}
$$
This result is known and the proof of other cases is similar.

Let $v_1=0$ or $v_2=0$.  Thus
the MS \er{VSE1} transforms to the scalar  NLS equation, since
$\|v_1\|, \|v_2\|$ are constants of motion.

\no ii) From \er{V23xx} we have \er{gb1}. Let $\cR\in \mR_2$. Then \er{uzs2} gives
$\t_1=e^{i\l}$ and then $\gb_1=0$.
\BBox

\no {\bf Proof of Corollary \ref{T7}.}
i) Let $v_2=pv_1$ for some constant $p\in \C\sm \{0\}$. Then
due to Lemma \ref{Tcom}
%we have $c_{12}=pc_{11}$. Then due to i) we have
$p$  is a constant of motion. From $v_2=pv_1$ we have
\[
\lb{yv1}
|v|^2=r_j^2|v_j|^2, \qqq  j=1,2,\qq           r_1^2=1+r^2,\qq r_2={r_1\/r}.
\]
where $r=|p|$. Thus we obtain the two equations for $y_j=r_jv_j$:
\[
\lb{yv2}
\begin{aligned}
i\dot v_j=-v_{j}''+ 2|v|^2v_j=-v_{j}''+ 2r_j^2|v_j|^2v_j\  \Leftrightarrow \ i\dot y_j=-y_j''+ 2|y_j|^2y_j,
\end{aligned}
\]
which yields  $y_2={p\/|p|}y_1$.

ii) Let $p=re^{i\vp}$ for some $r>0$. Using identities \er{yv1}, \er{yv2}  we construct the solution  $v=(v_1,v_2)\in \mH_o$ of the MS.
\BBox

\subsection{Entire functions}
We recall some well known facts about entire functions (see
\cite{Ks88}). An entire function $f(\l)$ is said to be of
$exponential$ $ type$ if there is a constant $\a$ such that
$|f(\l)|\leq $ const. $e^{\a |\l|}$ everywhere. The function $f$ is
said to belong to the Cartwright class $E_{Cart}(\gt_o),$ if $f(\l)$
is entire, of exponential type, and the following conditions hold
true:
$$
\int _{\R}{(\log |f(t)|)_+dt\/ 1+t^2}<\iy  ,\ \
\gt_o=\gt_{\pm}(f)>0, \qq {\rm where}\ \ \ \gt_{\pm}(f):= \lim
\sup_{s\to \pm\iy} {\log |f(is)|\/|s|},
$$
and $(x)_+=\max \{0, x\}$ for $x\in \R$.
We need the known result about the Hadamard factorization.
Let $f\in E_{Cart}(\gt_o)$ and denote by $(k_n)_{n=1}^{\iy} $ the
sequence of its zeros $\neq 0$ (counted with multiplicity), so
arranged that $0<|k_1|\leq |k_2|\leq \dots$. Then $f$ has the
Hadamard factorization
\[
\lb{2.7} \tes
 f(\l)=\l^mC\lim_{r\to +\iy}\prod_{|k_n|\leq
r}\lt(1-{\l\/k_n}\rt),\ \ \ \ C={f^{(m)}(0)\/m!},
\]
for some integer $m$, where the product converges uniformly in every
bounded disc and
\[
\lb{2.8}
 \sum_1^\iy {|\Im k_n|\/|k_n|^2}<\iy .
\]
Hence we obtain
\[
\lb{2.9}
 {f'(\l)\/ f(\l)}=i+{m\/ \l}+\lim_{r\to \iy}\sum_{|k_n|\leq
r}{1\/\l-k_n}
\]
uniformly on compact subsets of $\C\sm\big(\{0\}\cup\bigcup
\{k_n\}\big)$.  Denote by $\cN(r,f)$ the number of zeros of $f$
having modulus $\leq r$, each zero being counted according to its
multiplicity.

We recall the Levinson Theorem: {\it Let an entire
function $f\in E_{Cart}(\gt_o)$. Then}
\[
\lb{LeT} \cN (r,f)=\gt_o{2\/ \pi }r+o(r)\qq \as \qqq r\to \iy .
\]

%\newpage
 \begin{theorem}
\lb{Tdis} Let $v\in \mH$. Then $\vr\in E_{Cart}(2)$ and if $\l\in
\gS_1$, then there exists a multiplier $\t(\l)$ such that
$|\t(\l)|=e^{q(\l)}$, where $q(\l)>0$, and
 the following estimates hold true:
\[
\lb{esr1}  {(\sh^2q)\/4}  (\ch q-1)^2   \le |\gD(\l)|\le {\sh^2(2q)\/16}\qqq
\forall \ \ \l\in \gS_1,
\]
\[
\lb{esr2}
\begin{aligned}
{1\/\pi}\int_{\gS_1}|\gD(\l)|^{1\/6}d\l\ge 4^{1\/3}
Q_o={4^{5\/6}\/3}\|v\|^2,
\end{aligned}
\]
\[
\lb{esr3}
\begin{aligned}
 \tes {1\/\pi}\int_{\gS_1}\gD(\l)^{1\/2}d\l\le
{\ch(2\|v\|)\/2} Q_o.
\end{aligned}
\]
\end{theorem}

\no {\bf Proof.} In the proof of Theorem  \ref{Trho} we have proved
that there exists a multiplier $\t(\l)$ such that
$|\t(\l)|=e^{q(\l)}$, where $q(\l)>0$. From the identity  \er{rqmg}
we obtain for some $a\in [0,1]$
\[
\lb{rgap}
\begin{aligned}
& \tes 64|\gD(\l)|=16(\sh^2{q}) (\ch q-a)^2 16(\sh^2{q}) \ch^2 q\le
4\sh^2(2q).
\end{aligned}
\]
From the identity  \er{rqmg} we again obtain
$$
64|\gD(\l)|=16(\sh^2{q}) (\ch q-a)^2\ge 16(\sh^2{q}) (\ch q-1)^2\ge 4q^6
$$
and  jointly with the identity \er{T23-3} this implies
$$
{1\/\pi}\int_{\gS_1}|\gD(\l)|^{1\/6}d\l\ge
{4^{1\/3}\/\pi}\int_{\gS_1}q(\l)d\l=4^{1\/3} Q_o={4^{5\/6}\/3}\|v\|^2.
$$
From \er{esr1} and the simple estimate $\sh^2t\le t^2\ch^2t, t>0$ we
obtain
$$
64|\gD(\l)|\le 4 \sh^2(2q) \le 16 q^2\ch^2(2 q)\le 16 q^2\ch^2 (2
\|v\|)
$$
and this jointly with the identity \er{T23-3} implies
$$
{1\/\pi}\int_{\gS_1}|\gD(\l)|^{1\/2}d\l\le
{\ch (2 \|v\|)\/2}\int_{\gS_1}q(\l){d\l\/\pi}={\ch (2 \|v\|)\/2} Q_o.
$$
\BBox

\section {Appendix \lb{Sec5}}
\setcounter{equation}{0}

\no  Define the domain $G_n=G_n(1)=\{0< t_n<...<t_1< 1\}\ss \R^n$.

\begin{lemma} \lb{TA1}
Let $h_1,...,h_n\in L^2(0,1)$ and $u_n=t_1-t_2+\dots -(-1)^nt_n$ for
some $n\ge 3$. Then
\[
\lb{TA-1} \int_0^1dt\int_0^t e^{i2\l(t-s)}h_1(t)h_2(s)ds = {i\/2\l
}\lt(\int_0^1h_1(t)h_2(t)dt+o(1)\rt),
\]
\[
\lb{TA-2} \int_0^1dt\int_0^t
e^{-i2\l(t-s)}h_1(t)h_2(s)ds={o(e^{2\n})\/|\l|},
\]
\[
\lb{71} \int_0^1dt\int_0^t\cos \l (1-2t+2s)h_1(t)h_2(s)ds ={i\cos
\l\/2\l }\lt(\int_0^1h_1(t)h_2(t)dt+o(1)\rt),
\]
\[
\lb{TA-4} \int_{G_n} e^{\pm i\l (1-2u_n)}\prod_1^n h_j(t_j)dt
={o(e^{\n})\/|\l|},
\]
\[
\lb{ab2} \rt|\int_{G_n} e^{\pm i\l (1- 2 u_n)}\prod_1^n h_1(t_j) dt
\rt|\le {e^{\n}\/(4\n)^{{n\/4}}}{\|h_1\|^{n}\/n!},
\]
as $\l \in \mD_r$ and  $|\l|\to \iy$ for any fixed $r>0$, where
$\n=\Im \l$.
\end{lemma}

\no {\bf Proof.} We show \er{ab2}, since other estimtates were proved in \cite{K10}.  The integral $A_n^\pm:=\int_{G_n} e^{\pm i\l (1- 2
u_n)}\prod_1^n h(t_j) dt$  has an estimate
\[
\lb{ee1a} |A_n^\pm (\l)|^2\le \!\! e^{\mp 2\n}\rt(\int_{G_n}\!\!
e^{\pm 2\n u_n}\prod_1^n |h(t_j)| dt\rt)^{2} \le e^{\mp 2\n} f_n^\pm
\ {\|h\|^{n}\/n!},\qq f_n^\pm =\int_{G_n}\!\! e^{\pm 4\n u_n}dt,
\]
since $\int_{G_n}\!\! \prod_1^n |h(t_j)|^2 dt={\|h\|^{2n}\/n!}$. We
estimate the integral $f_n$. Direct computation gives
\[
\begin{aligned}
f_1^-={1\/4\n},\qq f_1^+={e^{4\n}\/4\n},\qq f_2^-\le {1\/4\n},\qq
f_2^+\le {e^{4\n}\/(4\n)^2}, \qq f_3^-\le {1\/(4\n)^2},\qq f_3^+\le
{e^{4\n}\/(4\n)^2}.
\end{aligned}
\]
Consider $n\ge 4$.  Below we need simple estimates
\[
\lb{n1}
\begin{aligned}
  \int_0^{t_{n-1}} e^{-4\n t_{n}}dt_{n}\le {1\/4\n},
 \end{aligned}
\]
\[
\lb{n2}
\begin{aligned}
 \int_0^{t_{n-2}} dt_{n-1}\int_0^{t_{n-1}} e^{4\n (-t_{n-1}
+t_{n})}dt_{n}\le \int_0^{t_{n-2}} dt_{n-1} e^{-4\n t_{n-1}}(e^{4\n
t_{n-1}}-1){dt_{n-1}\/4\n}\le {1\/4\n}.
\end{aligned}
\]
Consider $f_n^+$.  For even $n\ge 4$ from \er{n1} and after this
from \er{n2} we obtain
$$
f_n^+=\int_{G_n} e^{4\l(t_1-t_2+\dots+t_{n-1}-t_{n})}dt\le
{f_{n-1}^+\/4\n} \le {1\/(4\n)^{{n-2\/2}+1}}\int_0^{1} e^{4\n
t_1}dt_1 \le {e^{4\n}\/(4\n)^{{n\/2}+1}},
$$
and for  $n$ odd,  using \er{n2} and  we obtain
$$
f_n^+=\int_{G_n} e^{4\n(t_1-t_2+\dots-t_{n-1}+t_{n})}dt\le
{f_{n-2}^+\/4\n} \le {1\/(4\n)^{{n-1\/2}}}\int_0^{1} e^{4\n t_1}dt_1
\le {e^{4\n}\/(4\n)^{{n+1\/2}}}.
$$
Consider $f_n^-$. From  \er{n2} we obtain for even $n\ge 4$:
$$
f_n^-=\int_{G_n} e^{4\n(-t_1+t_2+\dots-t_{n-1}+t_{n})}dt\le
{f_{n-2}^-\/4\n} \le {1\/(4\n)^{{n\/2}}},
$$
and from \er{n1} and after this from \er{n2} for odd $n$:
$$
f_n^-=\int_{G_n} e^{4\n(-t_1+t_2+\dots+t_{n-1}-t_{n})}dt\le
{f_{n-1}^-\/4\n} \le {1\/(4\n)^{{n+1\/2}}}.
$$
Collecting all these integrals we obtain $ f_n^+ \le
{e^{4y}\/(4\n)^{{n+1\/2}}}$ and $ f_n^- \le {1\/(4\n)^{{n\/2}}} $
and jointly with \er{ee1} this gives \er{ab2}. \BBox

In the next lemma we recall   the asymptotic estimates of the
following integral
$$
\begin{aligned}
F_n^\pm(x)=\int_0^xe^{\pm i\l _nt}p(t)dt,\ \ \ \ G_n^{\pm
}(x)=\int_0^xe^{\mp i\l t}q(t)F_n^{\pm }(t)dt,\ \ \ \
p,q\in L^2(0,2\pi ),
\end{aligned}
$$
where $(x,\l_n,n)\in [0,2\pi ]\ts \C \ts \Z$.
For $p\in L^2(0,2\pi )$ we have the Fourier series $ p(x)=\sum
p_je^{ijx}$.

\begin{lemma}
\label{Ta2} Let  $C_0=4e^\pi$ and  $|z_n- \pi n|\le {1\/4}$. Then
\[
|F_n^{\pm}(x)|\le C_0u_{\pm n},\ \ \ |G_n^{\pm}(x)|\le C_0(u_{\pm
n}(q)u_{\pm n}(p)+v_{\pm n}^2+ w_{\pm n}^2),
\]
\[
u_{\pm }(p)=(p_{\pm n}(p))_{n\in \Z}, \ \ v_{\pm }=(v_{\pm n})_{n\in
\Z},\qq  w_{\pm }=(w_{\pm n})\in \ell^d(\Z), \qq \forall \ \ d>2,
\]
where
$$
u_{n}(p)= \sum {|p_j|\/ |j \pm n+i|},\ \ v_{n}^2=\sum {|q_jp_{-j}|\/
|j \pm n+i|},\ \ w_{n}^2= \sum {|p_jq_{s}|\/ |j\pm n+i||j+s+i|},\ \
n\in \Z ,
$$

\end{lemma}

\begin{lemma}
\label{Ta3} Let a function $f=2ie^{i\l}\sin\l$ define on the set  $\l\in \O_{\d,r}$ given by
$$
\O_{\d,r}=\{\l\in \ol\C_+: \Im \l\in [0,r],\  \Re \l\in [0,\pi/2],  \ |\l|\ge \d\},\ r>0, \ {1\/2}>\d>0.
$$
 Then
 \[
 \lb{aa1}
\d e^{-\d}<|f(\l)|<2 \qqq \forall \ l\in \O_{\d,r}.
\]
\end{lemma}
\no {\bf Proof.} The boundary of the set $\O_\d$ has the form
$
\pa \O_\d=\cup_o^4 I_j$, where
$$
\tes
I_o=\{\l=\d e^{i\vp}, \vp\in [0,{\pi\/2}]\}, \ \ I_1=[i\d, ir], \  \ I_2=[ir,ir+{\pi\/2}],\ \
I_3=[{\pi\/2}, {\pi\/2}+ir],\ \ I_4=[\d,{\pi\/2}].
$$
We have for all $\l\in \O_\d$:
\[
|f(\l)|=|e^{2i\l}-1|\le 2. 
\]
and $f=2ie^{i\l}\sin\l=e^{2i\l}-1$
\[
\begin{aligned}
|f(\l)|=|e^{2i\l}-1|\in [1-e^{-2\d},1-e^{-2r}],  \qq \l\in I_1,
\\
|f(\l)|=|e^{2i\l}-1|\in [1-e^{-2r},1+e^{-2r}],   \qq \l\in I_2,
\\
|f(\l)|=|e^{2i\l}-1|\in [2,1+e^{-2r}],   \qq \l\in I_3,
\\
|f(\l)|=2|\sin\l|\in [2\sin\d,2], \qq \l\in I_4.
\end{aligned}
\]
For $\l\in I_o$ we have $|f(\l)|=2e^{-\n}|\sin\l|$, where due to $\d\le {1\/2}$ we obtain
\[
\begin{aligned}
|\sin\l|=|\sum_{n=2j+1\ge 1}(-1)^{j} {\l^n\/n!}|\ge \d(1-{\d^2\/6}(1+\d^2+..))=\d(1-{\d^2\/6(1-\d^2})\ge {\d\/2}.
\end{aligned}
\]
This yields $|f(\l)|\ge e^{-\d}\d$ for $\l\in I_o$.
Collecting all estimates we get \er{aa1}.
\BBox

\footnotesize\footnotesize
  \no {\bf Acknowledgments.}
  E. K. was supported by the RSF grant  No.
19-71-30002.

%\newpage

\end{document}